\documentclass[aps,prb,10pt,twocolumn,superscriptaddress,longbibliography,nobalancelastpage]{revtex4-1}

\usepackage{graphicx}
\usepackage{graphics}
\usepackage{amsmath}
\usepackage{amssymb}
\usepackage{amsfonts}
\usepackage{dsfont}
\usepackage{braket}
\usepackage{color}
\usepackage{braket,slashed}
\usepackage[mathscr]{euscript}
\usepackage{hyperref}
\usepackage{subfigure}
\usepackage{xfrac}
\usepackage{bm}
\usepackage{kantlipsum}
\usepackage{enumitem}
\usepackage{tikz}
\usepackage{geometry}
\usepackage{framed}

\allowdisplaybreaks[1]

\newcommand{\code}[1]{\texttt{#1}}

\renewcommand{\dag}{^\dagger}

\newcommand{\e}{\ensuremath{\mathrm{e}}}

\newcommand{\vect}[1]{\ensuremath{\boldsymbol{#1}}}

\newcommand{\diagram}[2]{\,\vcenter{\hbox{\includegraphics[scale=0.3,page=#2]{./Diagrams/#1.pdf}}}\,}
\newcommand{\diagramsmall}[2]{\,\vcenter{\hbox{\includegraphics[scale=0.25,page=#2]{./Diagrams/#1.pdf}}}\,}
\newcommand{\rots}[0]{\text{rotations}}

\begin{document}

\title{Simulating excitation spectra with projected entangled-pair states}

\author{Laurens Vanderstraeten}
\email{laurens.vanderstraeten@ugent.be}
\affiliation{Department of Physics and Astronomy, University of Ghent, Krijgslaan 281, 9000 Gent, Belgium}

\author{Jutho Haegeman}
\affiliation{Department of Physics and Astronomy, University of Ghent, Krijgslaan 281, 9000 Gent, Belgium}

\author{Frank Verstraete}
\affiliation{Department of Physics and Astronomy, University of Ghent, Krijgslaan 281, 9000 Gent, Belgium}
\affiliation{Vienna Center for Quantum Science and Technology, Faculty of Physics, University of Vienna, Boltzmanngasse 5, 1090 Vienna, Austria}

\begin{abstract}
We develop and benchmark a technique for simulating excitation spectra of generic two-dimensional quantum lattice systems using the framework of projected entangled-pair states (PEPS). The technique relies on a variational ansatz for capturing quasiparticle excitations on top of a PEPS ground state. Our method perfectly captures the quasiparticle dispersion relation of the square-lattice transverse-field Ising model, and reproduces the spin-wave velocity and the spin-wave anomaly in the square-lattice Heisenberg model with high precision.
\end{abstract}

\maketitle

Quantum phases of matter are commonly characterized by the order of the ground states that are realized in many-body systems at zero temperature. Yet, the most interesting manifestations, both from the theoretical and the experimental perspective, of a given quantum phase are more related to the low-energy excitations on top of the ground state. In the case of strongly-correlated quantum phases these excitations typically have a collective nature, in the sense that they cannot be adiabatically connected to the excitations of a free system. Therefore, perturbative expansions or mean-field approximations do not allow for an accurate description, and more advanced conceptual and numerical tools have to be devised to understand their properties. This is especially true for one- and two-dimensional quantum systems, where strong quantum correlations often lead to collective excitations with fractionalized quantum numbers and anyonic statistics.
\par Tensor networks have emerged as a natural language for describing quantum phases of matter. It has been realized that these phases are characterized by the entanglement structure of their low-energy states, and that this structure can be captured by the class of tensor-network states \cite{Verstraete2008, Orus2013}. In one dimension, this yields the well-kown class of matrix product states \cite{Schollwock2011} (MPS) describing ground states of generic spin chains, as well as a powerful formalism for capturing quasiparticle excitations on top of these strongly-correlated background states \cite{Haegeman2012}. In this formalism, elementary excitations of the fully-interacting quantum Hamiltonian \cite{Haegeman2013} are naturally interpreted as interacting particles (including magnons, triplons spinons, chargeons, holons, etc. \cite{Zauner-Stauber2018}) and a two-particle S matrix can be defined and computed \cite{Vanderstraeten2015b}. For two-dimensional systems, the class of projected entangled-pair states (PEPS) \cite{Verstraete2004} provides the correct parametrization of ground states, and PEPS constructions for describing quasiparticles such as spinons \cite{Poilblanc2013} and anyons \cite{Schuch2010, Bultinck2017} have been proposed.
\par Yet, these constructions of quasiparticles against a PEPS background have been used numerically on fine-tuned wavefunctions only \cite{Haegeman2014, Iqbal2018}, and have not been applied variationally to characterize the spectrum of a given model Hamiltonian. Indeed, PEPS simulations have been restricted to the study of ground-state properties of ordered quantum magnets \cite{Corboz2011, Corboz2014a, Corboz2014b, Niesen2017}, (chiral) spin liquids \cite{Liao2017, Poilblanc2017a, Chen2018} \cite{}, and strongly-correlated electrons \cite{Corboz2014b,Zheng2017}, but, thanks to considerable progress in MPS techniques \cite{Vanderstraeten2015, Fishman2017}, in recent years a second generation of PEPS algorithms has started to form. The variational optimization of PEPS ground-state approximations has been made possible \cite{Corboz2016, Vanderstraeten2016}, which allows to obtain higher accuracy on ground-state energies and order parameters. This, in turn, has made it possible to develop accurate extrapolation techniques \cite{Rader2018, Corboz2018}. In addition, quantities such as momentum-resolved structure factors or the energy variance can be computed with high accuracy \cite{Vanderstraeten2016}.
\par These techniques now make it possible to go beyond ground-state properties, and to simulate the excitation spectrum of a given Hamiltonian with PEPS. A first step was taken in Ref.~\onlinecite{Vanderstraeten2015}, but limited to the special case of frustration-free Hamiltonians. In this work, we generalize this framework to simulate the excitation spectrum of generic Hamiltonians on two-dimensional lattices in the thermodynamic limit directly, and benchmark this method on the transverse-field Ising model and the square-lattice Heisenberg model.

\noindent\emph{The algorithm}--- %
The PEPS simulation of the excitation spectrum of a given model Hamiltonian $H=\sum_ih_i$ consists of two steps. In a first step, we find an optimal approximation for the model's ground state in terms of a PEPS tensor $A$. The ground-state wavefunction can be written down as
\begin{equation}
\ket{\Psi(A)} = \diagram{main}{1},
\end{equation}
which represents the contraction of an infinite network of the same five-leg tensor $A$,
\begin{equation}
A_{u,r,l,d}^s = \diagram{main}{2}\;.
\end{equation}
The index $s$ corresponds to the physical spin at each site, and the four virtual legs carry the correlations through the system; the dimension of these virtual legs is the so-called bond dimension $D$ of the PEPS, and serves as a control parameter in all PEPS simulations. Different algorithms were developed for finding an optimal PEPS representation with a given bond dimension in a variational way. In this work, we have used gradient-search methods \cite{Vanderstraeten2016} that variationally optimize the energy density directly in the thermodynamic limit.
\par In a next step, we build elementary excitations on top of this ground state. For these excitations we consider the variational ansatz
\begin{equation} \label{eq:ansatz}
\ket{\Phi_{\vec{q}}(B)} = \sum_{\vec{m}\in\mathcal{L}} \e^{i\vec{q}\cdot\vec{m}} \diagram{main}{3}.
\end{equation}
Here a new tensor $B$ is introduced at one site in the network (the round tensor), and a superposition is taken with momentum $\vec{q}=(q_x,q_y)$, which, because we work directly in the thermodynamic limit, ranges through the full Brillouin zone. Because the perturbation also acts on the virtual degrees of freedom in the PEPS, it can create a dressed object against a correlated background; in that sense, it can be interpreted as the PEPS generalization of the Feynman-Bijl ansatz \cite{Feynman1954} or single-mode approximation \cite{Arovas1988}.
\par In the ansatz wavefunction all variational freedom is contained within the tensor $B$. Moreover, as the wavefunction is clearly linear in the tensor $B$, the ansatz defines a linear subspace. The inner product within this subspace is found by computing the overlap between two states,
\begin{multline} \label{eq:effnorm}
\braket{\Phi_{\vec{q}\,'}(B') | \Phi_{\vec{q}}(B)} \\= 4\pi^2\delta^{(2)}(\vec{q}-\vec{q}\,')   (\vect{B}')\dag \vect{N}_{\vec{q}} \vect{B},
\end{multline}
where bold symbols denote the vectorized versions of the corresponding tensors. Up to a $\delta$ function normalization for the momenta, the inner product is determined by the effective norm matrix $\vect{N}_{\vec{q}}$. As was shown in Ref.~\onlinecite{Vanderstraeten2015}, some choices for the tensor $B$ give rise to a zero state in Eq.~\eqref{eq:ansatz}. Also, we want to confine our variational space to states that are locally orthogonal to the ground state. In the end, these restrictions lead to a basis $\vect{P}$ for the reduced subspace such that the projected norm matrix $\vect{P}\dag \vect{N}_{\vec{q}} \vect{P}$ is full rank.
\par Evaluating the overlap between two excited-state wavefunctions [Eq.~\eqref{eq:effnorm}], and a fortiori the computation of the effective norm matrix, requires an involved contraction. Indeed, this overlap reduces to a sum over all relative positions between the tensor $B$ in the ket vector and the tensor $B'$ in the bra. All these terms have the form of a two-point function
\begin{equation}
\diagramsmall{main}{7},
\end{equation}
where we have taken a top view of the double-layer (bra and ket) tensor-network diagram, the simple circle represents the contraction of a ground-state tensor $A$ and its conjugate, and the crosses denote where the $B$ tensor in the ket layer and the $\bar{B}'$ tensor in the bra layer are situated. It should be clear that these terms are determined by the correlations in the ground state, and in Ref.~\onlinecite{Vanderstraeten2015} a contraction scheme was developed that provides an effective channel environment for reducing the above diagram to an essentially one-dimensional contraction. In fact, these channel environments enable the full summation of all relative positions of the tensors in an efficient way, and, therefore, allow for an evaluation of $\vect{N}_{\vec{q}}$. The accuracy of this channel construction is controlled by the environment bond dimension $\chi$.
\par The variational optimization of the $B$ tensor now requires minimizing the excitation energy. Because of the linearity of the variational subspace, this amounts to solving the generalized eigenvalue problem
\begin{equation} \label{eq:eig}
 \left( \vect{P}\dag \vect{H}_{\vec{q}} \vect{P} \right) \vect{x}_{\vec{q}} = \omega(\vec{q}) \left( \vect{P}\dag \vect{N}_{\vec{q}} \vect{P} \right) \vect{x}_{\vec{q}}, 
\end{equation}
for the smallest-real eigenvalue. Here we have introduced the effective energy matrix
\begin{multline} \label{eq:Heff}
\bra{\Phi_{\vec{q}\,'}(B')} \tilde{H} \ket{ \Phi_{\vec{q}}(B)} \\= 4\pi^2\delta^{(2)}(\vec{q}-\vec{q}\,') (\vect{B}')\dag \vect{H}_{\vec{q}} \vect{B}.
\end{multline}
The eigenvalue $\omega(\vec{q})$ is the approximate excitation energy at momentum $\vec{q}$, whereas the state $\ket{\Phi_{\vec{q}}(B)}$ with $\vect{B}=\vect{P}\vect{x}_{\vec{q}}$, approximates the excited-state wavefunction. The renormalized Hamiltonian $\tilde{H}$ is obtained by subtracting the extensive ground-state energy, i.e. $\tilde{H}=\sum_i(h_i-\braket{h_i}_0)$, such that we find non-extensive excitation energies above the ground state.
\par The evaluation of the matrix elements of $\vect{H}_{\vec{q}}$ [Eq.~\eqref{eq:Heff}] involves the summation of three-point functions, but can, through a strategic positioning of channel environments, again be reduced to a number of one-dimensional contractions. We refer to the supplemental material for a full exposition of all tensor-network diagrams. The hardest diagrams, and therefore the computational complexity of implementing the generalized eigenvalue problem [Eq.~\eqref{eq:eig}], scales as $\mathcal{O}(D^6\chi^3+D^8\chi^2)$.

\noindent\emph{Benchmark results}--- %
As a first example we test our method on the transverse-field Ising model, defined by the Hamiltonian
\begin{equation}
H_{\text{ising}} = - \sum_{\braket{ij}} S_i^z S_j^z + \lambda \sum_i S_i^x.
\end{equation}
This model exhibits a quantum phase transition at $\lambda_c\approx3.044$ from a symmetry-broken phase ($\lambda<\lambda_c$) to a polarized phase ($\lambda>\lambda_c$). In Ref.~\onlinecite{Vanderstraeten2016} it was shown that the ground state of this model can be accurately approximated as a PEPS across the phase transition.
\par
\begin{figure}
\includegraphics[width=0.9\columnwidth]{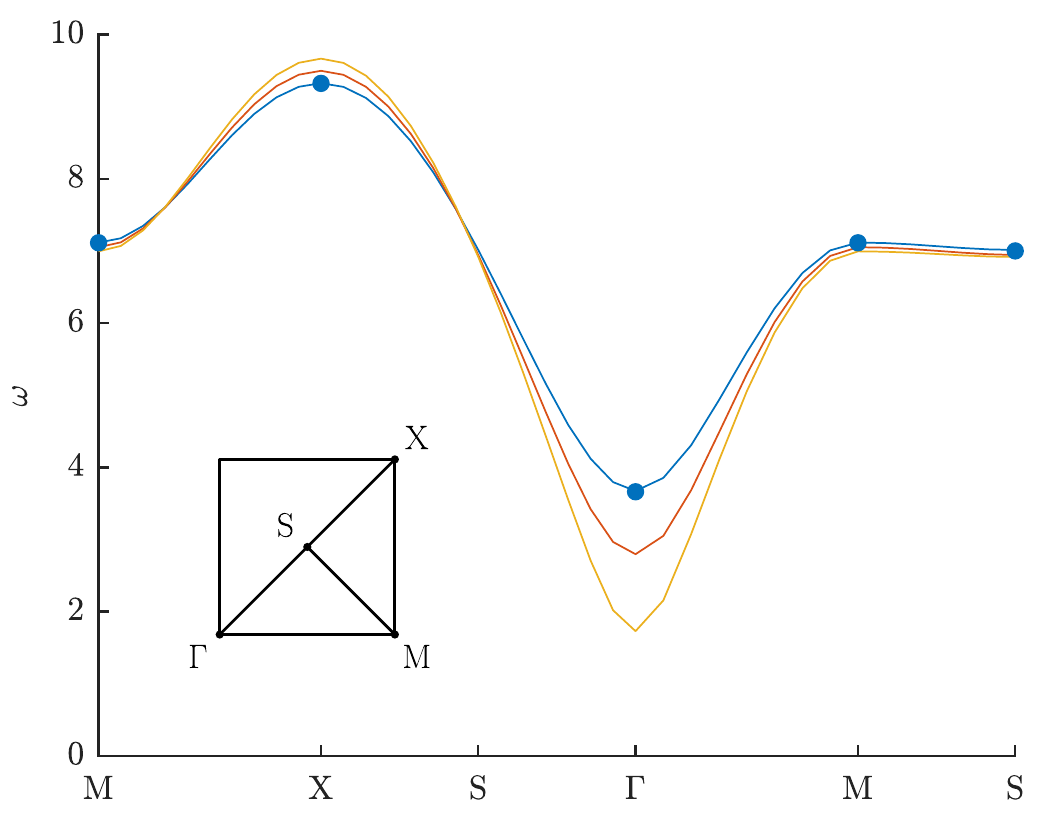}
\caption{The quasiparticle dispersion for the square-lattice transverse-field Ising model at $\lambda=2.5$ (blue), $\lambda=2.7$ (red) and $\lambda=2.9$ (yellow), as computed from PEPS with bond dimension $D=3$ and environment bond dimension $\chi=50$. The blue dots are series-expansion data points extracted from Ref.~\onlinecite{Oitmaa2006}.}
\label{fig:ising}
\end{figure}

\begin{figure}
\includegraphics[width=0.9\columnwidth]{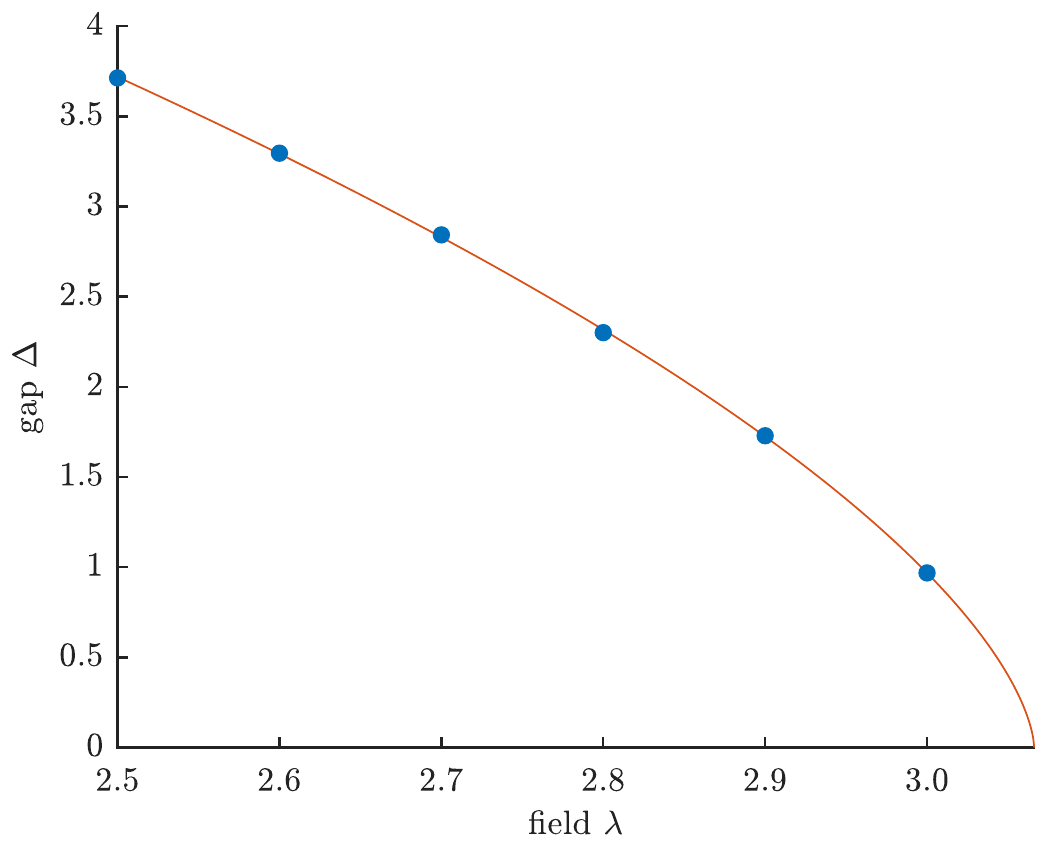}
\caption{The gap of the Ising model as a function of the field $\lambda$, with a fit $\Delta\propto\left|\lambda-\lambda_c\right|^\nu$; we find $\lambda_c\approx3.066$ and $\nu\approx0.627$. We have used $D=3$ and $\chi=50$.}
\label{fig:gap}
\end{figure}

 We can now build excitations on top of this ground state, and compute the quasiparticle spectrum. In Fig.~\ref{fig:ising} our results are plotted for the quasiparticle dispersion relation at three different values of the field; for one value of the field we compare with results from series expansions\cite{Oitmaa2006}. The dispersion clearly reaches a minimum at momentum $\vec{q}=(0,0)$, yielding the value of the gap. The phase transition is signalled by the vanishing of the excitation gap; in fact, close to the phase transition, the gap should obey a scaling relation with a critical exponent $\nu\approx0.6230$\cite{Oitmaa2006}. In Fig.~\ref{fig:gap} we have plotted the gap as a function of the field, showing that the power-law scaling is reproduced over a large region. We find estimates for the critical exponent $\nu$ and the critical field $\lambda$ within a percent precision.

\par A more challenging and interesting benchmark is provided by the Heisenberg model, defined by the Hamiltonian
\begin{equation}
H_{\text{heis}} = \sum_{\braket{ij}} S_i^xS_j^x - S_i^yS_j^y - S_i^zS_j^z.
\end{equation}
Here we have introduced the minus signs through a sublattice rotation of the original Heisenberg Hamiltonian, such that the staggered magnetization is mapped to a uniform one. Therefore, we can approximate the ground state as a uniform PEPS with a one-site unit cell \cite{Vanderstraeten2016}.

\begin{figure}
\includegraphics[width=0.9\columnwidth]{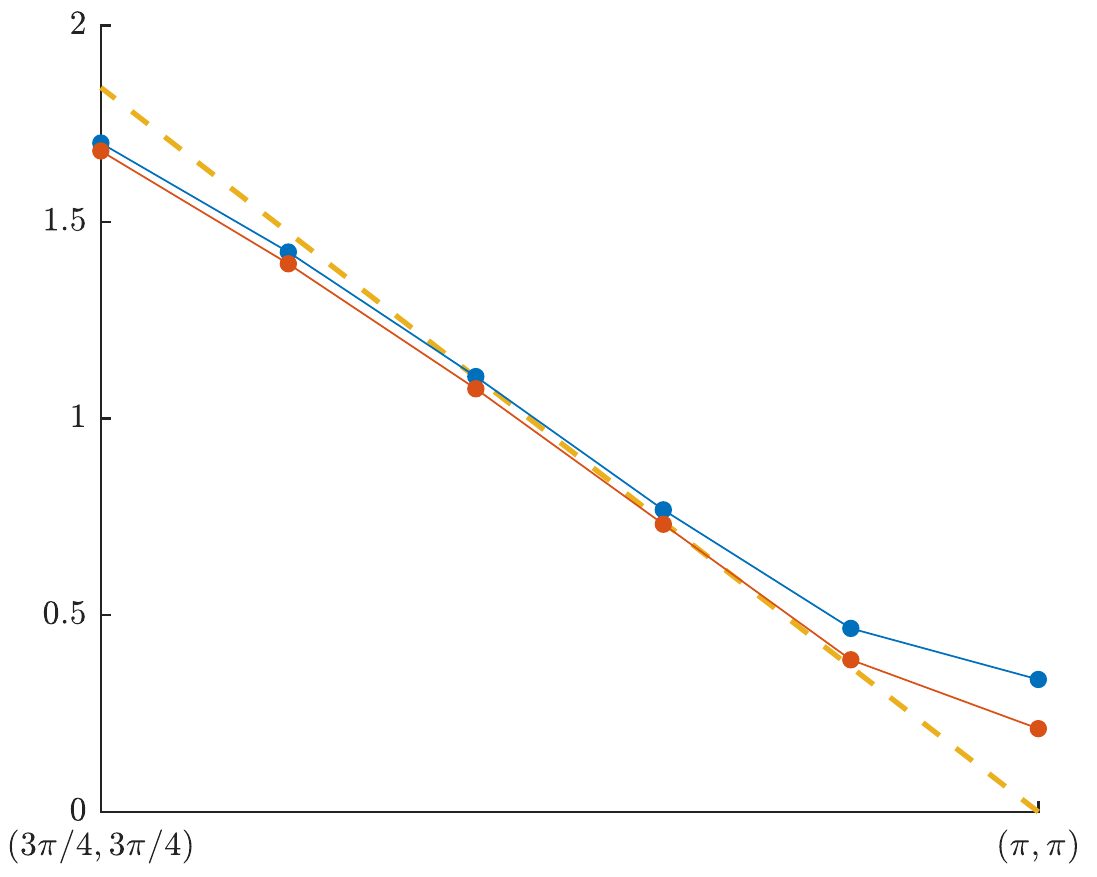}
\caption{The dispersion relation of the Heisenberg model approaching the gapless point $(\pi,0)$ with bond dimension $D=3$ (blue) and $D=4$ (red) and environment bond dimension up to $\chi=100$, as compared to the linear dispersion relation with $v_s\approx 1.65847$ \cite{Sen2015} (yellow). Clearly, the finite bond dimension of the PEPS induces an artificial gap, which grows smaller as $D$ increases. If we estimate the spin-wave velocity as the slope at the inflection point in the $D=4$ curve, we find $v_s\approx1.638$.}
\label{fig:xpoint}
\end{figure}

\par Since the ground state breaks a continuous symmetry, the spectrum exhibits a gapless Goldstone mode. The slope of the dispersion relation around the gapless point or spin-wave velocity $v_s$ is a crucial quantity in any low-energy field theory for this model \cite{Chakravarty1989, Chubukov1994} and is directly accessible in e.g. neutron-scattering experiments. In Fig.~\ref{fig:xpoint} we have plotted the dispersion relation in a small portion of the Brillouin zone where the gap is expected to close. Because the finite bond dimension induces a finite correlation length in the ground state \cite{Rader2018, Corboz2018}, the dispersion relation exhibits an artificial gap. This effect is clearly seen in our results, but is diminished as the bond dimension increases. Besides this effect, we reproduce the linear dispersion and we obtain an estimate for the spin-wave velocity that is close to the Monte-Carlo estimate.

\begin{table}[t]
	\begin{tabular}{ | p{2cm} |p{2cm} | p{2cm} |}
	\hline 
	& $(\pi/2,\pi/2)$ &  $(\pi,0)$  \\ \hline
	QMC \cite{Sandvik2001} & 2.4085 & 2.13 \\ \hline
    pCUT \cite{Powalski2015} & 2.375 & 2.2 \\ \hline
	ED \cite{Luscher2009} & 2.4144 & 2.2281 \\ \hline
	DMRG & 2.40  & 2.06-2.07 \\ \hline
	$D=4$ &  2.39 & 2.19 \\ \hline
	\end{tabular}
	\caption{The estimates for the excitation energies at wavevectors $\vec{q}=(\pi/2,\pi/2)$ and $\vec{q}=(\pi,0)$ as computed by some state-of-the-art numerical techniques, and compared to our PEPS result at $D=4$.}
	\label{tab:values}
\end{table}

\par
\begin{figure}
\includegraphics[width=0.9\columnwidth]{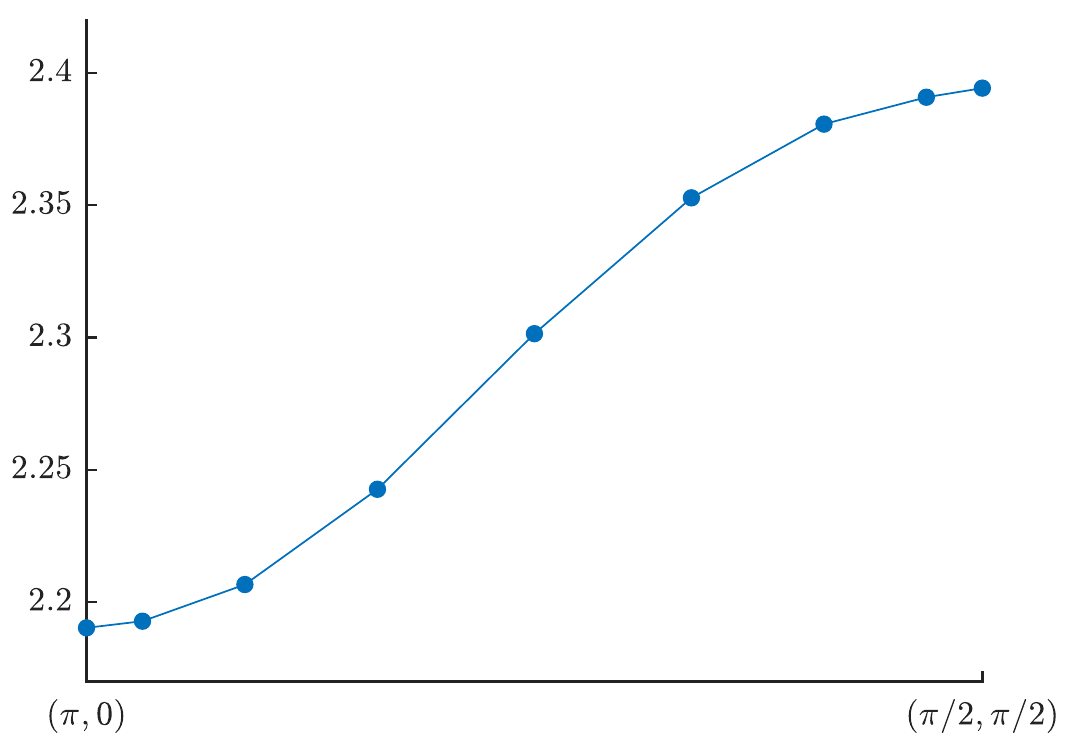}
\caption{The dispersion relation of the Heisenberg model between the points $(\pi,0)$ and $(\pi/2,\pi/2)$, computed with PEPS at $D=4$ and $\chi=100$.}
\label{fig:dip}
\end{figure}

 A more interesting feature of the spectrum is the shape of the dispersion relation at higher energies. Linear spin-wave theory predicts that the excitation gap is constant on the line between the wavevectors $(\pi,0)$ and $(\pi/2,\pi/2)$, but various numerical approaches \cite{Singh1995, Sandvik2001, Zheng2005, Luscher2009, Powalski2015, Shao2017, Powalski2018} and experimental measurements \cite{DallaPiazza2015} have shown that the excitation energy is suppressed at $(\pi,0)$ and elevated at $(\pi/2,\pi/2)$ as compared to the spin-wave result. The physical origin of this discrepancy has been argued to follow from spinon deconfinement around the wavevector $(\pi,0)$. In Fig.~\ref{fig:dip} we have plotted our results for the dispersion in this region, showing a significant dip in the dispersion. In Table \ref{tab:values} we compare our excitation energies to some other numerical results available in the literature. As to the physical origin of the spin-wave anomaly, we should note that the fact that our quasiparticle ansatz can accurately reproduce the excitation energy at $(\pi,0)$ suggests that this state is quite far from a deconfined two-spinon state.

\noindent\emph{Outlook}--- %
In this work we have presented a PEPS method for simulating excitation spectra of generic Hamiltonians in two dimensions, and have benchmarked the method on the Ising and Heisenberg models on the square lattice. The method can be readily extended to other lattice structures and PEPS ground states with larger unit cells. Although we have applied it to models that exhibit a local order parameter, our method can be applied equally well to more exotic quantum phases such as spin liquids and systems with topological order. In particular, the PEPS framework allows to directly target the fractionalized quasiparticles in these systems.
\par On the technical side many improvements of the method can be made. The conditioning of the generalized eigenvalue problem is, due to small eigenvalues in the effective norm matrix $\vect{N}_{\vec{q}}$, a source of large errors, and, therefore, a well-chosen preconditioner is required. Also, the implementation of $\vect{H}_{\vec{q}}$ requires the contraction of a large number of different diagrams and is therefore computationally demanding; better contraction schemes can speed up the simulations considerably. This, in turn, would make higher values of the PEPS bond dimension feasible, which would enable the use of extrapolation techniques for e.g. the excitation energies. Finally, the exploitation of symmetries in the PEPS representation of the ground state as well as the quasiparticle excitations will lead to much more efficient simulations.
\par In the context of matrix product states, this method of describing excitation spectra falls within the set of so-called tangent-space methods \cite{Vanderstraeten2018}. This numerical framework starts from the idea that the class of uniform MPS constitutes a manifold that describes ground states of spin chains, but realizes that the low-energy dynamics around such a ground state is contained within the tangent space on this manifold. In this work, we show that these tangent-space methods can also be applied to the PEPS manifold, and we expect that this will prove extremely useful in the simulation of the low-energy dynamics of two-dimensional quantum matter.

\noindent\emph{Acknowledgements}--- %
The authors would like to thank Kai Schmidt and Ruben Verresen for sharing their data. This work is supported by the Research Foundation Flanders, ERC grants QUTE (647905) and ERQUAF (715861), and the EU grant SIQS.​

\bibliography{./bibliography}

\begin{thebibliography}{42}%
\makeatletter
\providecommand \@ifxundefined [1]{%
 \@ifx{#1\undefined}
}%
\providecommand \@ifnum [1]{%
 \ifnum #1\expandafter \@firstoftwo
 \else \expandafter \@secondoftwo
 \fi
}%
\providecommand \@ifx [1]{%
 \ifx #1\expandafter \@firstoftwo
 \else \expandafter \@secondoftwo
 \fi
}%
\providecommand \natexlab [1]{#1}%
\providecommand \enquote  [1]{``#1''}%
\providecommand \bibnamefont  [1]{#1}%
\providecommand \bibfnamefont [1]{#1}%
\providecommand \citenamefont [1]{#1}%
\providecommand \href@noop [0]{\@secondoftwo}%
\providecommand \href [0]{\begingroup \@sanitize@url \@href}%
\providecommand \@href[1]{\@@startlink{#1}\@@href}%
\providecommand \@@href[1]{\endgroup#1\@@endlink}%
\providecommand \@sanitize@url [0]{\catcode `\\12\catcode `\$12\catcode
  `\&12\catcode `\#12\catcode `\^12\catcode `\_12\catcode `\%12\relax}%
\providecommand \@@startlink[1]{}%
\providecommand \@@endlink[0]{}%
\providecommand \url  [0]{\begingroup\@sanitize@url \@url }%
\providecommand \@url [1]{\endgroup\@href {#1}{\urlprefix }}%
\providecommand \urlprefix  [0]{URL }%
\providecommand \Eprint [0]{\href }%
\providecommand \doibase [0]{http://dx.doi.org/}%
\providecommand \selectlanguage [0]{\@gobble}%
\providecommand \bibinfo  [0]{\@secondoftwo}%
\providecommand \bibfield  [0]{\@secondoftwo}%
\providecommand \translation [1]{[#1]}%
\providecommand \BibitemOpen [0]{}%
\providecommand \bibitemStop [0]{}%
\providecommand \bibitemNoStop [0]{.\EOS\space}%
\providecommand \EOS [0]{\spacefactor3000\relax}%
\providecommand \BibitemShut  [1]{\csname bibitem#1\endcsname}%
\let\auto@bib@innerbib\@empty
\bibitem [{\citenamefont {Verstraete}\ \emph {et~al.}(2008)\citenamefont
  {Verstraete}, \citenamefont {Murg},\ and\ \citenamefont
  {Cirac}}]{Verstraete2008}%
  \BibitemOpen
  \bibfield  {author} {\bibinfo {author} {\bibfnamefont {F.}~\bibnamefont
  {Verstraete}}, \bibinfo {author} {\bibfnamefont {V.}~\bibnamefont {Murg}}, \
  and\ \bibinfo {author} {\bibfnamefont {J.~I.}\ \bibnamefont {Cirac}},\
  }\bibfield  {title} {\enquote {\bibinfo {title} {{Matrix product states,
  projected entangled pair states, and variational renormalization group
  methods for quantum spin systems}},}\ }\href {\doibase
  10.1080/14789940801912366} {\bibfield  {journal} {\bibinfo  {journal}
  {Advances in Physics}\ }\textbf {\bibinfo {volume} {57}},\ \bibinfo {pages}
  {143--224} (\bibinfo {year} {2008})}\BibitemShut {NoStop}%
\bibitem [{\citenamefont {Or{\'{u}}s}(2014)}]{Orus2013}%
  \BibitemOpen
  \bibfield  {author} {\bibinfo {author} {\bibfnamefont {R.}~\bibnamefont
  {Or{\'{u}}s}},\ }\bibfield  {title} {\enquote {\bibinfo {title} {{A practical
  introduction to tensor networks: Matrix product states and projected
  entangled pair states}},}\ }\href {\doibase 10.1016/j.aop.2014.06.013}
  {\bibfield  {journal} {\bibinfo  {journal} {Annals of Physics}\ }\textbf
  {\bibinfo {volume} {349}},\ \bibinfo {pages} {117--158} (\bibinfo {year}
  {2014})}\BibitemShut {NoStop}%
\bibitem [{\citenamefont {Schollw{\"{o}}ck}(2011)}]{Schollwock2011}%
  \BibitemOpen
  \bibfield  {author} {\bibinfo {author} {\bibfnamefont {U.}~\bibnamefont
  {Schollw{\"{o}}ck}},\ }\bibfield  {title} {\enquote {\bibinfo {title} {{The
  density-matrix renormalization group in the age of matrix product states}},}\
  }\href {\doibase 10.1016/j.aop.2010.09.012} {\bibfield  {journal} {\bibinfo
  {journal} {Annals of Physics}\ }\textbf {\bibinfo {volume} {326}},\ \bibinfo
  {pages} {96--192} (\bibinfo {year} {2011})}\BibitemShut {NoStop}%
\bibitem [{\citenamefont {Haegeman}\ \emph {et~al.}(2012)\citenamefont
  {Haegeman}, \citenamefont {Pirvu}, \citenamefont {Weir}, \citenamefont
  {Cirac}, \citenamefont {Osborne}, \citenamefont {Verschelde},\ and\
  \citenamefont {Verstraete}}]{Haegeman2012}%
  \BibitemOpen
  \bibfield  {author} {\bibinfo {author} {\bibfnamefont {J.}~\bibnamefont
  {Haegeman}}, \bibinfo {author} {\bibfnamefont {B.}~\bibnamefont {Pirvu}},
  \bibinfo {author} {\bibfnamefont {D.~J.}\ \bibnamefont {Weir}}, \bibinfo
  {author} {\bibfnamefont {J.~I.}\ \bibnamefont {Cirac}}, \bibinfo {author}
  {\bibfnamefont {T.~J.}\ \bibnamefont {Osborne}}, \bibinfo {author}
  {\bibfnamefont {H.}~\bibnamefont {Verschelde}}, \ and\ \bibinfo {author}
  {\bibfnamefont {F.}~\bibnamefont {Verstraete}},\ }\bibfield  {title}
  {\enquote {\bibinfo {title} {Variational matrix product ansatz for dispersion
  relations},}\ }\href {\doibase 10.1103/PhysRevB.85.100408} {\bibfield
  {journal} {\bibinfo  {journal} {Physical Review B}\ }\textbf {\bibinfo
  {volume} {85}},\ \bibinfo {pages} {100408} (\bibinfo {year}
  {2012})}\BibitemShut {NoStop}%
\bibitem [{\citenamefont {Haegeman}\ \emph {et~al.}(2013)\citenamefont
  {Haegeman}, \citenamefont {Michalakis}, \citenamefont {Nachtergaele},
  \citenamefont {Osborne}, \citenamefont {Schuch},\ and\ \citenamefont
  {Verstraete}}]{Haegeman2013}%
  \BibitemOpen
  \bibfield  {author} {\bibinfo {author} {\bibfnamefont {J.}~\bibnamefont
  {Haegeman}}, \bibinfo {author} {\bibfnamefont {S.}~\bibnamefont
  {Michalakis}}, \bibinfo {author} {\bibfnamefont {B.}~\bibnamefont
  {Nachtergaele}}, \bibinfo {author} {\bibfnamefont {T.~J.}\ \bibnamefont
  {Osborne}}, \bibinfo {author} {\bibfnamefont {N.}~\bibnamefont {Schuch}}, \
  and\ \bibinfo {author} {\bibfnamefont {F.}~\bibnamefont {Verstraete}},\
  }\bibfield  {title} {\enquote {\bibinfo {title} {Elementary excitations in
  gapped quantum spin systems},}\ }\href {\doibase
  10.1103/PhysRevLett.111.080401} {\bibfield  {journal} {\bibinfo  {journal}
  {Physical Review Letters}\ }\textbf {\bibinfo {volume} {111}},\ \bibinfo
  {pages} {080401} (\bibinfo {year} {2013})}\BibitemShut {NoStop}%
\bibitem [{\citenamefont {Zauner-Stauber}\ \emph {et~al.}(2018)\citenamefont
  {Zauner-Stauber}, \citenamefont {Vanderstraeten}, \citenamefont {Haegeman},
  \citenamefont {McCulloch},\ and\ \citenamefont
  {Verstraete}}]{Zauner-Stauber2018}%
  \BibitemOpen
  \bibfield  {author} {\bibinfo {author} {\bibfnamefont {V.}~\bibnamefont
  {Zauner-Stauber}}, \bibinfo {author} {\bibfnamefont {L.}~\bibnamefont
  {Vanderstraeten}}, \bibinfo {author} {\bibfnamefont {J.}~\bibnamefont
  {Haegeman}}, \bibinfo {author} {\bibfnamefont {I.~P.}\ \bibnamefont
  {McCulloch}}, \ and\ \bibinfo {author} {\bibfnamefont {F.}~\bibnamefont
  {Verstraete}},\ }\bibfield  {title} {\enquote {\bibinfo {title} {Topological
  nature of spinons and holons: Elementary excitations from matrix product
  states with conserved symmetries},}\ }\href {\doibase
  10.1103/PhysRevB.97.235155} {\bibfield  {journal} {\bibinfo  {journal}
  {Physical Review B}\ }\textbf {\bibinfo {volume} {97}},\ \bibinfo {pages}
  {235155} (\bibinfo {year} {2018})}\BibitemShut {NoStop}%
\bibitem [{\citenamefont {Vanderstraeten}\ \emph
  {et~al.}(2015{\natexlab{a}})\citenamefont {Vanderstraeten}, \citenamefont
  {Verstraete},\ and\ \citenamefont {Haegeman}}]{Vanderstraeten2015b}%
  \BibitemOpen
  \bibfield  {author} {\bibinfo {author} {\bibfnamefont {L.}~\bibnamefont
  {Vanderstraeten}}, \bibinfo {author} {\bibfnamefont {F.}~\bibnamefont
  {Verstraete}}, \ and\ \bibinfo {author} {\bibfnamefont {J.}~\bibnamefont
  {Haegeman}},\ }\bibfield  {title} {\enquote {\bibinfo {title} {Scattering
  particles in quantum spin chains},}\ }\href {\doibase
  10.1103/PhysRevB.92.125136} {\bibfield  {journal} {\bibinfo  {journal}
  {Physical Review B}\ }\textbf {\bibinfo {volume} {92}},\ \bibinfo {pages}
  {125136} (\bibinfo {year} {2015}{\natexlab{a}})}\BibitemShut {NoStop}%
\bibitem [{\citenamefont {Verstraete}\ and\ \citenamefont
  {Cirac}(2004)}]{Verstraete2004}%
  \BibitemOpen
  \bibfield  {author} {\bibinfo {author} {\bibfnamefont {F.}~\bibnamefont
  {Verstraete}}\ and\ \bibinfo {author} {\bibfnamefont {J.~I.}\ \bibnamefont
  {Cirac}},\ }\bibfield  {title} {\enquote {\bibinfo {title} {{Renormalization
  algorithms for Quantum-Many Body Systems in two and higher dimensions}},}\
  }\href@noop {} {\bibfield  {journal} {\bibinfo  {journal} {arXiv}\ }
  (\bibinfo {year} {2004})},\ \Eprint {http://arxiv.org/abs/cond-mat/0407066}
  {cond-mat/0407066} \BibitemShut {NoStop}%
\bibitem [{\citenamefont {Poilblanc}\ and\ \citenamefont
  {Schuch}(2013)}]{Poilblanc2013}%
  \BibitemOpen
  \bibfield  {author} {\bibinfo {author} {\bibfnamefont {D.}~\bibnamefont
  {Poilblanc}}\ and\ \bibinfo {author} {\bibfnamefont {N.}~\bibnamefont
  {Schuch}},\ }\bibfield  {title} {\enquote {\bibinfo {title} {Simplex $z_{2}$
  spin liquids on the kagome lattice with projected entangled pair states:
  Spinon and vison coherence lengths, topological entropy, and gapless edge
  modes},}\ }\href {\doibase 10.1103/PhysRevB.87.140407} {\bibfield  {journal}
  {\bibinfo  {journal} {Physical Review B}\ }\textbf {\bibinfo {volume} {87}},\
  \bibinfo {pages} {140407} (\bibinfo {year} {2013})}\BibitemShut {NoStop}%
\bibitem [{\citenamefont {Schuch}\ \emph {et~al.}(2010)\citenamefont {Schuch},
  \citenamefont {Cirac},\ and\ \citenamefont
  {P\'{e}rez-Garc\'{i}a}}]{Schuch2010}%
  \BibitemOpen
  \bibfield  {author} {\bibinfo {author} {\bibfnamefont {N.}~\bibnamefont
  {Schuch}}, \bibinfo {author} {\bibfnamefont {J.~I.}\ \bibnamefont {Cirac}}, \
  and\ \bibinfo {author} {\bibfnamefont {D.}~\bibnamefont
  {P\'{e}rez-Garc\'{i}a}},\ }\bibfield  {title} {\enquote {\bibinfo {title}
  {Peps as ground states: Degeneracy and topology},}\ }\href {\doibase
  https://doi.org/10.1016/j.aop.2010.05.008} {\bibfield  {journal} {\bibinfo
  {journal} {Annals of Physics}\ }\textbf {\bibinfo {volume} {325}},\ \bibinfo
  {pages} {2153 -- 2192} (\bibinfo {year} {2010})}\BibitemShut {NoStop}%
\bibitem [{\citenamefont {Bultinck}\ \emph {et~al.}(2017)\citenamefont
  {Bultinck}, \citenamefont {Mariën}, \citenamefont {Williamson},
  \citenamefont {\c{S}ahino\u{g}lu}, \citenamefont {Haegeman},\ and\
  \citenamefont {Verstraete}}]{Bultinck2017}%
  \BibitemOpen
  \bibfield  {author} {\bibinfo {author} {\bibfnamefont {N.}~\bibnamefont
  {Bultinck}}, \bibinfo {author} {\bibfnamefont {M.}~\bibnamefont {Mariën}},
  \bibinfo {author} {\bibfnamefont {D.J.}\ \bibnamefont {Williamson}}, \bibinfo
  {author} {\bibfnamefont {M.~B.}\ \bibnamefont {\c{S}ahino\u{g}lu}}, \bibinfo
  {author} {\bibfnamefont {J.}~\bibnamefont {Haegeman}}, \ and\ \bibinfo
  {author} {\bibfnamefont {F.}~\bibnamefont {Verstraete}},\ }\bibfield  {title}
  {\enquote {\bibinfo {title} {Anyons and matrix product operator algebras},}\
  }\href {\doibase https://doi.org/10.1016/j.aop.2017.01.004} {\bibfield
  {journal} {\bibinfo  {journal} {Annals of Physics}\ }\textbf {\bibinfo
  {volume} {378}},\ \bibinfo {pages} {183 -- 233} (\bibinfo {year}
  {2017})}\BibitemShut {NoStop}%
\bibitem [{\citenamefont {Haegeman}\ \emph {et~al.}(2015)\citenamefont
  {Haegeman}, \citenamefont {Zauner}, \citenamefont {Schuch},\ and\
  \citenamefont {Verstraete}}]{Haegeman2014}%
  \BibitemOpen
  \bibfield  {author} {\bibinfo {author} {\bibfnamefont {J.}~\bibnamefont
  {Haegeman}}, \bibinfo {author} {\bibfnamefont {V.}~\bibnamefont {Zauner}},
  \bibinfo {author} {\bibfnamefont {N.}~\bibnamefont {Schuch}}, \ and\ \bibinfo
  {author} {\bibfnamefont {F.}~\bibnamefont {Verstraete}},\ }\bibfield  {title}
  {\enquote {\bibinfo {title} {{Shadows of anyons and the entanglement
  structure of topological phases}},}\ }\href {\doibase 10.1038/ncomms9284}
  {\bibfield  {journal} {\bibinfo  {journal} {Nature Communications}\ }\textbf
  {\bibinfo {volume} {6}},\ \bibinfo {pages} {8284} (\bibinfo {year}
  {2015})}\BibitemShut {NoStop}%
\bibitem [{\citenamefont {Iqbal}\ \emph {et~al.}(2018)\citenamefont {Iqbal},
  \citenamefont {Duivenvoorden},\ and\ \citenamefont {Schuch}}]{Iqbal2018}%
  \BibitemOpen
  \bibfield  {author} {\bibinfo {author} {\bibfnamefont {M.}~\bibnamefont
  {Iqbal}}, \bibinfo {author} {\bibfnamefont {K.}~\bibnamefont
  {Duivenvoorden}}, \ and\ \bibinfo {author} {\bibfnamefont {N.}~\bibnamefont
  {Schuch}},\ }\bibfield  {title} {\enquote {\bibinfo {title} {Study of anyon
  condensation and topological phase transitions from a $z_{4}$ topological
  phase using the projected entangled pair states approach},}\ }\href {\doibase
  10.1103/PhysRevB.97.195124} {\bibfield  {journal} {\bibinfo  {journal}
  {Physical Review B}\ }\textbf {\bibinfo {volume} {97}},\ \bibinfo {pages}
  {195124} (\bibinfo {year} {2018})}\BibitemShut {NoStop}%
\bibitem [{\citenamefont {Corboz}\ \emph {et~al.}(2011)\citenamefont {Corboz},
  \citenamefont {L\"auchli}, \citenamefont {Penc}, \citenamefont {Troyer},\
  and\ \citenamefont {Mila}}]{Corboz2011}%
  \BibitemOpen
  \bibfield  {author} {\bibinfo {author} {\bibfnamefont {P.}~\bibnamefont
  {Corboz}}, \bibinfo {author} {\bibfnamefont {A.~M.}\ \bibnamefont
  {L\"auchli}}, \bibinfo {author} {\bibfnamefont {K.}~\bibnamefont {Penc}},
  \bibinfo {author} {\bibfnamefont {M.}~\bibnamefont {Troyer}}, \ and\ \bibinfo
  {author} {\bibfnamefont {F.}~\bibnamefont {Mila}},\ }\bibfield  {title}
  {\enquote {\bibinfo {title} {Simultaneous dimerization and su(4) symmetry
  breaking of 4-color fermions on the square lattice},}\ }\href {\doibase
  10.1103/PhysRevLett.107.215301} {\bibfield  {journal} {\bibinfo  {journal}
  {Physical Review Letters}\ }\textbf {\bibinfo {volume} {107}},\ \bibinfo
  {pages} {215301} (\bibinfo {year} {2011})}\BibitemShut {NoStop}%
\bibitem [{\citenamefont {Corboz}\ and\ \citenamefont
  {Mila}(2014)}]{Corboz2014a}%
  \BibitemOpen
  \bibfield  {author} {\bibinfo {author} {\bibfnamefont {P.}~\bibnamefont
  {Corboz}}\ and\ \bibinfo {author} {\bibfnamefont {F.}~\bibnamefont {Mila}},\
  }\bibfield  {title} {\enquote {\bibinfo {title} {Crystals of bound states in
  the magnetization plateaus of the shastry-sutherland model},}\ }\href
  {\doibase 10.1103/PhysRevLett.112.147203} {\bibfield  {journal} {\bibinfo
  {journal} {Physical Review Letters}\ }\textbf {\bibinfo {volume} {112}},\
  \bibinfo {pages} {147203} (\bibinfo {year} {2014})}\BibitemShut {NoStop}%
\bibitem [{\citenamefont {Corboz}\ \emph {et~al.}(2014)\citenamefont {Corboz},
  \citenamefont {Rice},\ and\ \citenamefont {Troyer}}]{Corboz2014b}%
  \BibitemOpen
  \bibfield  {author} {\bibinfo {author} {\bibfnamefont {P.}~\bibnamefont
  {Corboz}}, \bibinfo {author} {\bibfnamefont {T.~M.}\ \bibnamefont {Rice}}, \
  and\ \bibinfo {author} {\bibfnamefont {M.}~\bibnamefont {Troyer}},\
  }\bibfield  {title} {\enquote {\bibinfo {title} {Competing states in the
  $t$-$j$ model: Uniform $d$-wave state versus stripe state},}\ }\href
  {\doibase 10.1103/PhysRevLett.113.046402} {\bibfield  {journal} {\bibinfo
  {journal} {Physical Review Letters}\ }\textbf {\bibinfo {volume} {113}},\
  \bibinfo {pages} {046402} (\bibinfo {year} {2014})}\BibitemShut {NoStop}%
\bibitem [{\citenamefont {Niesen}\ and\ \citenamefont
  {Corboz}(2017)}]{Niesen2017}%
  \BibitemOpen
  \bibfield  {author} {\bibinfo {author} {\bibfnamefont {I.}~\bibnamefont
  {Niesen}}\ and\ \bibinfo {author} {\bibfnamefont {P.}~\bibnamefont
  {Corboz}},\ }\bibfield  {title} {\enquote {\bibinfo {title} {Emergent haldane
  phase in the $s=1$ bilinear-biquadratic heisenberg model on the square
  lattice},}\ }\href {\doibase 10.1103/PhysRevB.95.180404} {\bibfield
  {journal} {\bibinfo  {journal} {Physical Review B}\ }\textbf {\bibinfo
  {volume} {95}},\ \bibinfo {pages} {180404} (\bibinfo {year}
  {2017})}\BibitemShut {NoStop}%
\bibitem [{\citenamefont {Liao}\ \emph {et~al.}(2017)\citenamefont {Liao},
  \citenamefont {Xie}, \citenamefont {Chen}, \citenamefont {Liu}, \citenamefont
  {Xie}, \citenamefont {Huang}, \citenamefont {Normand},\ and\ \citenamefont
  {Xiang}}]{Liao2017}%
  \BibitemOpen
  \bibfield  {author} {\bibinfo {author} {\bibfnamefont {H.~J.}\ \bibnamefont
  {Liao}}, \bibinfo {author} {\bibfnamefont {Z.~Y.}\ \bibnamefont {Xie}},
  \bibinfo {author} {\bibfnamefont {J.}~\bibnamefont {Chen}}, \bibinfo {author}
  {\bibfnamefont {Z.~Y.}\ \bibnamefont {Liu}}, \bibinfo {author} {\bibfnamefont
  {H.~D.}\ \bibnamefont {Xie}}, \bibinfo {author} {\bibfnamefont {R.~Z.}\
  \bibnamefont {Huang}}, \bibinfo {author} {\bibfnamefont {B.}~\bibnamefont
  {Normand}}, \ and\ \bibinfo {author} {\bibfnamefont {T.}~\bibnamefont
  {Xiang}},\ }\bibfield  {title} {\enquote {\bibinfo {title} {Gapless
  spin-liquid ground state in the $s=1/2$ kagome antiferromagnet},}\ }\href
  {\doibase 10.1103/PhysRevLett.118.137202} {\bibfield  {journal} {\bibinfo
  {journal} {Physical Review Letters}\ }\textbf {\bibinfo {volume} {118}},\
  \bibinfo {pages} {137202} (\bibinfo {year} {2017})}\BibitemShut {NoStop}%
\bibitem [{\citenamefont {Poilblanc}\ and\ \citenamefont
  {Mambrini}(2017)}]{Poilblanc2017a}%
  \BibitemOpen
  \bibfield  {author} {\bibinfo {author} {\bibfnamefont {D.}~\bibnamefont
  {Poilblanc}}\ and\ \bibinfo {author} {\bibfnamefont {M.}~\bibnamefont
  {Mambrini}},\ }\bibfield  {title} {\enquote {\bibinfo {title} {Quantum
  critical phase with infinite projected entangled paired states},}\ }\href
  {\doibase 10.1103/PhysRevB.96.014414} {\bibfield  {journal} {\bibinfo
  {journal} {Physical Review B}\ }\textbf {\bibinfo {volume} {96}},\ \bibinfo
  {pages} {014414} (\bibinfo {year} {2017})}\BibitemShut {NoStop}%
\bibitem [{\citenamefont {Chen}\ \emph {et~al.}(2018)\citenamefont {Chen},
  \citenamefont {Vanderstraeten}, \citenamefont {Capponi},\ and\ \citenamefont
  {Poilblanc}}]{Chen2018}%
  \BibitemOpen
  \bibfield  {author} {\bibinfo {author} {\bibfnamefont {J.-Y.}\ \bibnamefont
  {Chen}}, \bibinfo {author} {\bibfnamefont {L.}~\bibnamefont
  {Vanderstraeten}}, \bibinfo {author} {\bibfnamefont {S.}~\bibnamefont
  {Capponi}}, \ and\ \bibinfo {author} {\bibfnamefont {D.}~\bibnamefont
  {Poilblanc}},\ }\bibfield  {title} {\enquote {\bibinfo {title} {Non-abelian
  chiral spin liquid in a quantum antiferromagnet revealed by an ipeps
  study},}\ }\href@noop {} {\bibfield  {journal} {\bibinfo  {journal} {arXiv}\
  } (\bibinfo {year} {2018})},\ \Eprint {http://arxiv.org/abs/1807.04385}
  {1807.04385} \BibitemShut {NoStop}%
\bibitem [{\citenamefont {Zheng}\ \emph {et~al.}(2017)\citenamefont {Zheng},
  \citenamefont {Chung}, \citenamefont {Corboz}, \citenamefont {Ehlers},
  \citenamefont {Qin}, \citenamefont {Noack}, \citenamefont {Shi},
  \citenamefont {White}, \citenamefont {Zhang},\ and\ \citenamefont
  {Chan}}]{Zheng2017}%
  \BibitemOpen
  \bibfield  {author} {\bibinfo {author} {\bibfnamefont {B.-X.}\ \bibnamefont
  {Zheng}}, \bibinfo {author} {\bibfnamefont {C.-M.}\ \bibnamefont {Chung}},
  \bibinfo {author} {\bibfnamefont {P.}~\bibnamefont {Corboz}}, \bibinfo
  {author} {\bibfnamefont {G.}~\bibnamefont {Ehlers}}, \bibinfo {author}
  {\bibfnamefont {M.-P.}\ \bibnamefont {Qin}}, \bibinfo {author} {\bibfnamefont
  {R.~M.}\ \bibnamefont {Noack}}, \bibinfo {author} {\bibfnamefont
  {H.}~\bibnamefont {Shi}}, \bibinfo {author} {\bibfnamefont {S.~R.}\
  \bibnamefont {White}}, \bibinfo {author} {\bibfnamefont {S.}~\bibnamefont
  {Zhang}}, \ and\ \bibinfo {author} {\bibfnamefont {G.~K.-L.}\ \bibnamefont
  {Chan}},\ }\bibfield  {title} {\enquote {\bibinfo {title} {Stripe order in
  the underdoped region of the two-dimensional hubbard model},}\ }\href
  {\doibase 10.1126/science.aam7127} {\bibfield  {journal} {\bibinfo  {journal}
  {Science}\ }\textbf {\bibinfo {volume} {358}},\ \bibinfo {pages} {1155--1160}
  (\bibinfo {year} {2017})}\BibitemShut {NoStop}%
\bibitem [{\citenamefont {Vanderstraeten}\ \emph
  {et~al.}(2015{\natexlab{b}})\citenamefont {Vanderstraeten}, \citenamefont
  {Mari\"en}, \citenamefont {Verstraete},\ and\ \citenamefont
  {Haegeman}}]{Vanderstraeten2015}%
  \BibitemOpen
  \bibfield  {author} {\bibinfo {author} {\bibfnamefont {L.}~\bibnamefont
  {Vanderstraeten}}, \bibinfo {author} {\bibfnamefont {M.}~\bibnamefont
  {Mari\"en}}, \bibinfo {author} {\bibfnamefont {F.}~\bibnamefont
  {Verstraete}}, \ and\ \bibinfo {author} {\bibfnamefont {J.}~\bibnamefont
  {Haegeman}},\ }\bibfield  {title} {\enquote {\bibinfo {title} {Excitations
  and the tangent space of projected entangled-pair states},}\ }\href {\doibase
  10.1103/PhysRevB.92.201111} {\bibfield  {journal} {\bibinfo  {journal}
  {Physical Review B}\ }\textbf {\bibinfo {volume} {92}},\ \bibinfo {pages}
  {201111} (\bibinfo {year} {2015}{\natexlab{b}})}\BibitemShut {NoStop}%
\bibitem [{\citenamefont {Fishman}\ \emph {et~al.}(2017)\citenamefont
  {Fishman}, \citenamefont {Vanderstraeten}, \citenamefont {Zauner-Stauber},
  \citenamefont {Haegeman},\ and\ \citenamefont {Verstraete}}]{Fishman2017}%
  \BibitemOpen
  \bibfield  {author} {\bibinfo {author} {\bibfnamefont {M.T.}\ \bibnamefont
  {Fishman}}, \bibinfo {author} {\bibfnamefont {L.}~\bibnamefont
  {Vanderstraeten}}, \bibinfo {author} {\bibfnamefont {V.}~\bibnamefont
  {Zauner-Stauber}}, \bibinfo {author} {\bibfnamefont {J.}~\bibnamefont
  {Haegeman}}, \ and\ \bibinfo {author} {\bibfnamefont {F.}~\bibnamefont
  {Verstraete}},\ }\bibfield  {title} {\enquote {\bibinfo {title} {Faster
  methods for contracting infinite 2d tensor networks},}\ }\href@noop {}
  {\bibfield  {journal} {\bibinfo  {journal} {arXiv}\ } (\bibinfo {year}
  {2017})},\ \Eprint {http://arxiv.org/abs/1711.05881} {1711.05881}
  \BibitemShut {NoStop}%
\bibitem [{\citenamefont {Corboz}(2016)}]{Corboz2016}%
  \BibitemOpen
  \bibfield  {author} {\bibinfo {author} {\bibfnamefont {P.}~\bibnamefont
  {Corboz}},\ }\bibfield  {title} {\enquote {\bibinfo {title} {Variational
  optimization with infinite projected entangled-pair states},}\ }\href
  {\doibase 10.1103/PhysRevB.94.035133} {\bibfield  {journal} {\bibinfo
  {journal} {Physical Review B}\ }\textbf {\bibinfo {volume} {94}},\ \bibinfo
  {pages} {035133} (\bibinfo {year} {2016})}\BibitemShut {NoStop}%
\bibitem [{\citenamefont {Vanderstraeten}\ \emph {et~al.}(2016)\citenamefont
  {Vanderstraeten}, \citenamefont {Haegeman}, \citenamefont {Corboz},\ and\
  \citenamefont {Verstraete}}]{Vanderstraeten2016}%
  \BibitemOpen
  \bibfield  {author} {\bibinfo {author} {\bibfnamefont {L.}~\bibnamefont
  {Vanderstraeten}}, \bibinfo {author} {\bibfnamefont {J.}~\bibnamefont
  {Haegeman}}, \bibinfo {author} {\bibfnamefont {P.}~\bibnamefont {Corboz}}, \
  and\ \bibinfo {author} {\bibfnamefont {F.}~\bibnamefont {Verstraete}},\
  }\bibfield  {title} {\enquote {\bibinfo {title} {Gradient methods for
  variational optimization of projected entangled-pair states},}\ }\href
  {\doibase 10.1103/PhysRevB.94.155123} {\bibfield  {journal} {\bibinfo
  {journal} {Physical Review B}\ }\textbf {\bibinfo {volume} {94}},\ \bibinfo
  {pages} {155123} (\bibinfo {year} {2016})}\BibitemShut {NoStop}%
\bibitem [{\citenamefont {Rader}\ and\ \citenamefont
  {L\"auchli}(2018)}]{Rader2018}%
  \BibitemOpen
  \bibfield  {author} {\bibinfo {author} {\bibfnamefont {M.}~\bibnamefont
  {Rader}}\ and\ \bibinfo {author} {\bibfnamefont {A.~M.}\ \bibnamefont
  {L\"auchli}},\ }\bibfield  {title} {\enquote {\bibinfo {title} {Finite
  correlation length scaling in lorentz-invariant gapless ipeps wave
  functions},}\ }\href {\doibase 10.1103/PhysRevX.8.031030} {\bibfield
  {journal} {\bibinfo  {journal} {Physical Review X}\ }\textbf {\bibinfo
  {volume} {8}},\ \bibinfo {pages} {031030} (\bibinfo {year}
  {2018})}\BibitemShut {NoStop}%
\bibitem [{\citenamefont {Corboz}\ \emph {et~al.}(2018)\citenamefont {Corboz},
  \citenamefont {Czarnik}, \citenamefont {Kapteijns},\ and\ \citenamefont
  {Tagliacozzo}}]{Corboz2018}%
  \BibitemOpen
  \bibfield  {author} {\bibinfo {author} {\bibfnamefont {P.}~\bibnamefont
  {Corboz}}, \bibinfo {author} {\bibfnamefont {P.}~\bibnamefont {Czarnik}},
  \bibinfo {author} {\bibfnamefont {G.}~\bibnamefont {Kapteijns}}, \ and\
  \bibinfo {author} {\bibfnamefont {L.}~\bibnamefont {Tagliacozzo}},\
  }\bibfield  {title} {\enquote {\bibinfo {title} {Finite correlation length
  scaling with infinite projected entangled-pair states},}\ }\href {\doibase
  10.1103/PhysRevX.8.031031} {\bibfield  {journal} {\bibinfo  {journal}
  {Physical Review X}\ }\textbf {\bibinfo {volume} {8}},\ \bibinfo {pages}
  {031031} (\bibinfo {year} {2018})}\BibitemShut {NoStop}%
\bibitem [{\citenamefont {Feynman}(1954)}]{Feynman1954}%
  \BibitemOpen
  \bibfield  {author} {\bibinfo {author} {\bibfnamefont {R.~P.}\ \bibnamefont
  {Feynman}},\ }\bibfield  {title} {\enquote {\bibinfo {title} {{Atomic Theory
  of the Two-Fluid Model of Liquid Helium}},}\ }\href {\doibase
  10.1103/PhysRev.94.262} {\bibfield  {journal} {\bibinfo  {journal} {Physical
  Review}\ }\textbf {\bibinfo {volume} {94}},\ \bibinfo {pages} {262--277}
  (\bibinfo {year} {1954})}\BibitemShut {NoStop}%
\bibitem [{\citenamefont {Arovas}\ \emph {et~al.}(1988)\citenamefont {Arovas},
  \citenamefont {Auerbach},\ and\ \citenamefont {Haldane}}]{Arovas1988}%
  \BibitemOpen
  \bibfield  {author} {\bibinfo {author} {\bibfnamefont {D.~P.}\ \bibnamefont
  {Arovas}}, \bibinfo {author} {\bibfnamefont {A.}~\bibnamefont {Auerbach}}, \
  and\ \bibinfo {author} {\bibfnamefont {F.~D.~M.}\ \bibnamefont {Haldane}},\
  }\bibfield  {title} {\enquote {\bibinfo {title} {{Extended Heisenberg models
  of antiferromagnetism: Analogies to the fractional quantum Hall effect}},}\
  }\href {http://link.aps.org/doi/10.1103/PhysRevLett.60.531} {\bibfield
  {journal} {\bibinfo  {journal} {Physical Review Letters}\ }\textbf {\bibinfo
  {volume} {60}},\ \bibinfo {pages} {531--534} (\bibinfo {year}
  {1988})}\BibitemShut {NoStop}%
\bibitem [{\citenamefont {Oitmaa}\ \emph {et~al.}(2006)\citenamefont {Oitmaa},
  \citenamefont {Hamer},\ and\ \citenamefont {Zheng}}]{Oitmaa2006}%
  \BibitemOpen
  \bibfield  {author} {\bibinfo {author} {\bibfnamefont {J.}~\bibnamefont
  {Oitmaa}}, \bibinfo {author} {\bibfnamefont {C.}~\bibnamefont {Hamer}}, \
  and\ \bibinfo {author} {\bibfnamefont {W.}~\bibnamefont {Zheng}},\
  }\href@noop {} {\emph {\bibinfo {title} {Series Expansion Methods for
  Strongly Interacting Lattice Models}}}\ (\bibinfo  {publisher} {Cambridge
  University Press},\ \bibinfo {year} {2006})\BibitemShut {NoStop}%
\bibitem [{\citenamefont {Sen}\ \emph {et~al.}(2015)\citenamefont {Sen},
  \citenamefont {Suwa},\ and\ \citenamefont {Sandvik}}]{Sen2015}%
  \BibitemOpen
  \bibfield  {author} {\bibinfo {author} {\bibfnamefont {A.}~\bibnamefont
  {Sen}}, \bibinfo {author} {\bibfnamefont {H.}~\bibnamefont {Suwa}}, \ and\
  \bibinfo {author} {\bibfnamefont {A.~W.}\ \bibnamefont {Sandvik}},\
  }\bibfield  {title} {\enquote {\bibinfo {title} {Velocity of excitations in
  ordered, disordered, and critical antiferromagnets},}\ }\href {\doibase
  10.1103/PhysRevB.92.195145} {\bibfield  {journal} {\bibinfo  {journal}
  {Physical Review B}\ }\textbf {\bibinfo {volume} {92}},\ \bibinfo {pages}
  {195145} (\bibinfo {year} {2015})}\BibitemShut {NoStop}%
\bibitem [{\citenamefont {Chakravarty}\ \emph {et~al.}(1989)\citenamefont
  {Chakravarty}, \citenamefont {Halperin},\ and\ \citenamefont
  {Nelson}}]{Chakravarty1989}%
  \BibitemOpen
  \bibfield  {author} {\bibinfo {author} {\bibfnamefont {S.}~\bibnamefont
  {Chakravarty}}, \bibinfo {author} {\bibfnamefont {B.~I.}\ \bibnamefont
  {Halperin}}, \ and\ \bibinfo {author} {\bibfnamefont {D.~R.}\ \bibnamefont
  {Nelson}},\ }\bibfield  {title} {\enquote {\bibinfo {title} {Two-dimensional
  quantum heisenberg antiferromagnet at low temperatures},}\ }\href {\doibase
  10.1103/PhysRevB.39.2344} {\bibfield  {journal} {\bibinfo  {journal}
  {Physical Review B}\ }\textbf {\bibinfo {volume} {39}},\ \bibinfo {pages}
  {2344--2371} (\bibinfo {year} {1989})}\BibitemShut {NoStop}%
\bibitem [{\citenamefont {Chubukov}\ \emph {et~al.}(1994)\citenamefont
  {Chubukov}, \citenamefont {Sachdev},\ and\ \citenamefont
  {Ye}}]{Chubukov1994}%
  \BibitemOpen
  \bibfield  {author} {\bibinfo {author} {\bibfnamefont {A.~V.}\ \bibnamefont
  {Chubukov}}, \bibinfo {author} {\bibfnamefont {S.}~\bibnamefont {Sachdev}}, \
  and\ \bibinfo {author} {\bibfnamefont {J.}~\bibnamefont {Ye}},\ }\bibfield
  {title} {\enquote {\bibinfo {title} {Theory of two-dimensional quantum
  heisenberg antiferromagnets with a nearly critical ground state},}\ }\href
  {\doibase 10.1103/PhysRevB.49.11919} {\bibfield  {journal} {\bibinfo
  {journal} {Physical Review B}\ }\textbf {\bibinfo {volume} {49}},\ \bibinfo
  {pages} {11919--11961} (\bibinfo {year} {1994})}\BibitemShut {NoStop}%
\bibitem [{\citenamefont {Sandvik}\ and\ \citenamefont
  {Singh}(2001)}]{Sandvik2001}%
  \BibitemOpen
  \bibfield  {author} {\bibinfo {author} {\bibfnamefont {A.~W.}\ \bibnamefont
  {Sandvik}}\ and\ \bibinfo {author} {\bibfnamefont {R.~R.~P.}\ \bibnamefont
  {Singh}},\ }\bibfield  {title} {\enquote {\bibinfo {title} {High-energy
  magnon dispersion and multimagnon continuum in the two-dimensional heisenberg
  antiferromagnet},}\ }\href {\doibase 10.1103/PhysRevLett.86.528} {\bibfield
  {journal} {\bibinfo  {journal} {Physical Review Letters}\ }\textbf {\bibinfo
  {volume} {86}},\ \bibinfo {pages} {528--531} (\bibinfo {year}
  {2001})}\BibitemShut {NoStop}%
\bibitem [{\citenamefont {Powalski}\ \emph {et~al.}(2015)\citenamefont
  {Powalski}, \citenamefont {Uhrig},\ and\ \citenamefont
  {Schmidt}}]{Powalski2015}%
  \BibitemOpen
  \bibfield  {author} {\bibinfo {author} {\bibfnamefont {M.}~\bibnamefont
  {Powalski}}, \bibinfo {author} {\bibfnamefont {G.~S.}\ \bibnamefont {Uhrig}},
  \ and\ \bibinfo {author} {\bibfnamefont {K.~P.}\ \bibnamefont {Schmidt}},\
  }\bibfield  {title} {\enquote {\bibinfo {title} {Roton minimum as a
  fingerprint of magnon-higgs scattering in ordered quantum
  antiferromagnets},}\ }\href {\doibase 10.1103/PhysRevLett.115.207202}
  {\bibfield  {journal} {\bibinfo  {journal} {Physical Review Letters}\
  }\textbf {\bibinfo {volume} {115}},\ \bibinfo {pages} {207202} (\bibinfo
  {year} {2015})}\BibitemShut {NoStop}%
\bibitem [{\citenamefont {L\"uscher}\ and\ \citenamefont
  {L\"auchli}(2009)}]{Luscher2009}%
  \BibitemOpen
  \bibfield  {author} {\bibinfo {author} {\bibfnamefont {A.}~\bibnamefont
  {L\"uscher}}\ and\ \bibinfo {author} {\bibfnamefont {A.~M.}\ \bibnamefont
  {L\"auchli}},\ }\bibfield  {title} {\enquote {\bibinfo {title} {Exact
  diagonalization study of the antiferromagnetic spin-$\frac{1}{2}$ heisenberg
  model on the square lattice in a magnetic field},}\ }\href {\doibase
  10.1103/PhysRevB.79.195102} {\bibfield  {journal} {\bibinfo  {journal}
  {Physical Review B}\ }\textbf {\bibinfo {volume} {79}},\ \bibinfo {pages}
  {195102} (\bibinfo {year} {2009})}\BibitemShut {NoStop}%
\bibitem [{\citenamefont {Singh}\ and\ \citenamefont
  {Gelfand}(1995)}]{Singh1995}%
  \BibitemOpen
  \bibfield  {author} {\bibinfo {author} {\bibfnamefont {R.~R.~P.}\
  \bibnamefont {Singh}}\ and\ \bibinfo {author} {\bibfnamefont {M.~P.}\
  \bibnamefont {Gelfand}},\ }\bibfield  {title} {\enquote {\bibinfo {title}
  {Spin-wave excitation spectra and spectral weights in square lattice
  antiferromagnets},}\ }\href {\doibase 10.1103/PhysRevB.52.R15695} {\bibfield
  {journal} {\bibinfo  {journal} {Physical Review B}\ }\textbf {\bibinfo
  {volume} {52}},\ \bibinfo {pages} {R15695--R15698} (\bibinfo {year}
  {1995})}\BibitemShut {NoStop}%
\bibitem [{\citenamefont {Zheng}\ \emph {et~al.}(2005)\citenamefont {Zheng},
  \citenamefont {Oitmaa},\ and\ \citenamefont {Hamer}}]{Zheng2005}%
  \BibitemOpen
  \bibfield  {author} {\bibinfo {author} {\bibfnamefont {W.}~\bibnamefont
  {Zheng}}, \bibinfo {author} {\bibfnamefont {J.}~\bibnamefont {Oitmaa}}, \
  and\ \bibinfo {author} {\bibfnamefont {C.~J.}\ \bibnamefont {Hamer}},\
  }\bibfield  {title} {\enquote {\bibinfo {title} {Series studies of the
  spin-$\frac{1}{2}$ heisenberg antiferromagnet at $t=0$: Magnon dispersion and
  structure factors},}\ }\href {\doibase 10.1103/PhysRevB.71.184440} {\bibfield
   {journal} {\bibinfo  {journal} {Physical Review B}\ }\textbf {\bibinfo
  {volume} {71}},\ \bibinfo {pages} {184440} (\bibinfo {year}
  {2005})}\BibitemShut {NoStop}%
\bibitem [{\citenamefont {Shao}\ \emph {et~al.}(2017)\citenamefont {Shao},
  \citenamefont {Qin}, \citenamefont {Capponi}, \citenamefont {Chesi},
  \citenamefont {Meng},\ and\ \citenamefont {Sandvik}}]{Shao2017}%
  \BibitemOpen
  \bibfield  {author} {\bibinfo {author} {\bibfnamefont {H.}~\bibnamefont
  {Shao}}, \bibinfo {author} {\bibfnamefont {Y.~Q.}\ \bibnamefont {Qin}},
  \bibinfo {author} {\bibfnamefont {S.}~\bibnamefont {Capponi}}, \bibinfo
  {author} {\bibfnamefont {S.}~\bibnamefont {Chesi}}, \bibinfo {author}
  {\bibfnamefont {Z.~Y.}\ \bibnamefont {Meng}}, \ and\ \bibinfo {author}
  {\bibfnamefont {A.~W.}\ \bibnamefont {Sandvik}},\ }\bibfield  {title}
  {\enquote {\bibinfo {title} {Nearly deconfined spinon excitations in the
  square-lattice spin-$1/2$ heisenberg antiferromagnet},}\ }\href {\doibase
  10.1103/PhysRevX.7.041072} {\bibfield  {journal} {\bibinfo  {journal}
  {Physical Review X}\ }\textbf {\bibinfo {volume} {7}},\ \bibinfo {pages}
  {041072} (\bibinfo {year} {2017})}\BibitemShut {NoStop}%
\bibitem [{\citenamefont {Powalski}\ \emph {et~al.}(2018)\citenamefont
  {Powalski}, \citenamefont {Schmidt},\ and\ \citenamefont
  {Uhrig}}]{Powalski2018}%
  \BibitemOpen
  \bibfield  {author} {\bibinfo {author} {\bibfnamefont {M.}~\bibnamefont
  {Powalski}}, \bibinfo {author} {\bibfnamefont {K.~P.}\ \bibnamefont
  {Schmidt}}, \ and\ \bibinfo {author} {\bibfnamefont {G.~S.}\ \bibnamefont
  {Uhrig}},\ }\bibfield  {title} {\enquote {\bibinfo {title} {{Mutually
  attracting spin waves in the square-lattice quantum antiferromagnet}},}\
  }\href {\doibase 10.21468/SciPostPhys.4.1.001} {\bibfield  {journal}
  {\bibinfo  {journal} {SciPost Phys.}\ }\textbf {\bibinfo {volume} {4}},\
  \bibinfo {pages} {001} (\bibinfo {year} {2018})}\BibitemShut {NoStop}%
\bibitem [{\citenamefont {Piazza}\ \emph {et~al.}(2015)\citenamefont {Piazza},
  \citenamefont {Mourigal}, \citenamefont {Christensen}, \citenamefont
  {Nilsen}, \citenamefont {Tregenna-Piggott}, \citenamefont {Perring},
  \citenamefont {Enderle}, \citenamefont {McMorrow}, \citenamefont {Ivanov},\
  and\ \citenamefont {R{\o}nnow}}]{DallaPiazza2015}%
  \BibitemOpen
  \bibfield  {author} {\bibinfo {author} {\bibfnamefont {B.~Dalla}\
  \bibnamefont {Piazza}}, \bibinfo {author} {\bibfnamefont {M.}~\bibnamefont
  {Mourigal}}, \bibinfo {author} {\bibfnamefont {N.~B.}\ \bibnamefont
  {Christensen}}, \bibinfo {author} {\bibfnamefont {G.~J.}\ \bibnamefont
  {Nilsen}}, \bibinfo {author} {\bibfnamefont {P.}~\bibnamefont
  {Tregenna-Piggott}}, \bibinfo {author} {\bibfnamefont {T.~G.}\ \bibnamefont
  {Perring}}, \bibinfo {author} {\bibfnamefont {M.}~\bibnamefont {Enderle}},
  \bibinfo {author} {\bibfnamefont {D.~F.}\ \bibnamefont {McMorrow}}, \bibinfo
  {author} {\bibfnamefont {D.~A.}\ \bibnamefont {Ivanov}}, \ and\ \bibinfo
  {author} {\bibfnamefont {H.~M.}\ \bibnamefont {R{\o}nnow}},\ }\bibfield
  {title} {\enquote {\bibinfo {title} {Fractional excitations in the
  square-lattice quantum antiferromagnet},}\ }\href {\doibase
  https://doi.org/10.1038/nphys3172} {\bibfield  {journal} {\bibinfo  {journal}
  {Nature Physics}\ }\textbf {\bibinfo {volume} {11}},\ \bibinfo {pages} {62}
  (\bibinfo {year} {2015})}\BibitemShut {NoStop}%
\bibitem [{\citenamefont {Vanderstraeten}\ \emph {et~al.}()\citenamefont
  {Vanderstraeten}, \citenamefont {Haegeman},\ and\ \citenamefont
  {Verstraete}}]{Vanderstraeten2018}%
  \BibitemOpen
  \bibfield  {author} {\bibinfo {author} {\bibfnamefont {L.}~\bibnamefont
  {Vanderstraeten}}, \bibinfo {author} {\bibfnamefont {J.}~\bibnamefont
  {Haegeman}}, \ and\ \bibinfo {author} {\bibfnamefont {F.}~\bibnamefont
  {Verstraete}},\ }\bibfield  {title} {\enquote {\bibinfo {title}
  {Tangent-space methods for uniform matrix product states},}\ }\href@noop {}
  {\bibinfo  {journal} {in preparation}\ }\BibitemShut {NoStop}%
\end{thebibliography}%

\newpage
\setcounter{equation}{0}
\appendix
\onecolumngrid

\section*{Appendix: Implementing the PEPS quasiparticle ansatz}

In this appendix we provide the details for implementing and optimizing the PEPS excitation ansatz. We work on the square lattice for which the ground state can be represented as a translation- and rotation-invariant PEPS, and we take the case of a translation invariant Hamiltonian with terms acting on a plaquette of four sites. We assume that we have an optimized ground-state approximation, for the details of how to variationally optimize the ground-state PEPS tensor we refer to Ref. \onlinecite{Vanderstraeten2016}. In order to make this appendix self-contained, we recapitulate the construction of effective channel environments \cite{Vanderstraeten2015} in a first section, and then explain the excitation ansatz in detail.

\subsection{Uniform PEPS and effective environments}

We assume that the ground state of the Hamiltonian is given by a PEPS,
\begin{equation}
\ket{\Psi(A)} = \diagram{peps}{1}
\end{equation}
where the wiggly lines correspond to the physical degrees of freedom. The tensors appearing in the PEPS are all copies of the same tensor $A$,
\begin{equation}
A_{u,r,d,l}^s = \diagram{peps}{2},
\end{equation}
and we assume that the tensor is rotation invariant,
\begin{equation}
A_{u,r,d,l}^s = A_{l,u,r,d}^s = A_{d,l,u,r}^s = A_{r,d,l,u}^s,
\end{equation}
so that the state $\ket{\Psi(A)}$ is too (everything below can be readily generalized to non-rotation invariant case).
\par The norm of a PEPS can be represented as
\begin{equation}
\braket{\Psi(A)|\Psi(A)} = \diagram{peps}{3},
\end{equation}
where the round tensor represents the contraction of an $A$ tensor with its complex conjugate along the physical dimension, and we have grouped two legs into one in each direction,
\begin{equation}
\diagram{peps}{4} = \sum_{s,s'} \delta_{s,s'} A_{u,r,d,l}^s \bar{A}_{u',r',d',l'}^{s'}.
\end{equation}
The contraction of this infinite tensor network is performed by finding the fixed point of the linear transfer matrix in the form of an MPS on the virtual level. This MPS obeys the fixed-point equation
\begin{equation}
\diagram{peps}{5} \propto \diagram{peps}{6} \;.
\end{equation}
Here, the MPS is represented by a triple of tensors $\{A_l,A_r,C\}$ which obey the conditions
\begin{equation}
\diagram{peps}{7} = \diagram{peps}{8} \;,
\end{equation}
and
\begin{equation}
\diagram{peps}{9} = \diagram{peps}{10}, \hspace{2cm} \diagram{peps}{11} = \diagram{peps}{12} \;.
\end{equation}
The `norm per site' $f=-\frac{1}{N}\log(\braket{\Psi(A)|\Psi(A)})$ corresponding to the PEPS is then defined as
\begin{equation}
f = - \log \lambda,
\end{equation}
where $\lambda$ is given as the leading eigenvalue of the following eigenvalue equations
\begin{equation}
\diagram{peps}{13} = \lambda \diagram{peps}{14}, \hspace{2cm} \diagram{peps}{15} = \lambda \diagram{peps}{16} \;
\end{equation}
The corresponding eigenvectors $G_l$ and $G_r$ are normalized such that
\begin{equation}
\frac{1}{\lambda} \diagram{peps}{17} = \diagram{peps}{18} = 1
\end{equation}
We can now rescale the PEPS tensor $A$ as
\begin{equation}
A \rightarrow A/\sqrt{\lambda},
\end{equation}
such that we have $f=0$, and all PEPS expectation values are well-defined in the thermodynamic limit.
\par In addition, we define a corner tensor $S$ which is the leading eigenvector of the equation
\begin{equation}
\diagram{peps}{30} \propto \diagram{peps}{31},
\end{equation}
which can be simplified to
\begin{equation}
\diagram{peps}{32} \propto \diagram{peps}{31}.
\end{equation}
\par In order to compute expectation values we find four environments $\{M_l,M_r,C\}_i$ which we use to determine the contraction of the norm of the PEPS as
\begin{equation}
\diagram{peps}{19}.
\end{equation}
The MPS tensors $\{M_l,M_r\}$ are given by $\{A_l,A_r\}$ but normalized such that the following eigenvalue equation has leading eigenvalue $\mu=1$:
\begin{equation}
\diagram{peps}{20} = \mu \diagram{peps}{21}.
\end{equation}
The PEPS can now be fully normalized by rescaling the corresponding eigenvector $\rho$ such that
\begin{equation}
\braket{\Psi(A)|\Psi(A)} = \diagram{peps}{22} = 1.
\end{equation}
We also define the fixed point in the other direction $\tilde{\rho}$ as
\begin{equation}
\diagram{peps}{23} = \mu \diagram{peps}{24}
\end{equation}
and normalize it such that
\begin{equation}
\diagram{peps}{25} = 1.
\end{equation}
The norm of the PEPS is then also given by
\begin{equation}
\braket{\Psi(A)|\Psi(A)} = \diagram{peps}{26} = 1
\end{equation}
\par We now define the corner environment
\begin{equation}
\diagram{peps}{27},
\end{equation}
where the corner matrix $S$ is normalized such that, again, the norm of the PEPS is one:
\begin{equation}
\braket{\Psi(A)|\Psi(A)} = \diagram{peps}{28} = \diagram{peps}{29}  = 1.
\end{equation}
\par We also define two-site channel fixed points as
\begin{equation}
\diagram{energy}{1} = \mu \diagram{energy}{2}.
\end{equation}
Because of the redefinition of $M_l$ and $M_r$, the eigenvalue $\mu$ should be (approximately) one. We also define the other fixed point, 
\begin{equation}
\diagram{energy}{3} = \mu \diagram{energy}{4}.
\end{equation}
Note that we overload the symbols for tensors, where its definition depends on the number of legs it has. The energy density expectation value is given by the two equivalent diagrams
\begin{equation}
e = \diagram{energy}{5} = \diagram{energy}{6},
\end{equation}
where we have introduced the dashed lines to indicate the four sites on which the Hamiltonian plaquette operator acts.

\subsection{The ansatz wavefunction}

We use the following ansatz for an excitation with momentum $(q_x,q_y)$
\begin{equation}
\ket{\Phi_{\vec{q}}(B)} = \sum_{k,l} \e^{iq_xk} \e^{iq_yl} \diagram{ansatz}{1}
\end{equation}
where the round tensor denotes a new tensor $B$,
\begin{equation}
B_{u,r,d,l}^s = \diagram{ansatz}{2},
\end{equation}
containing all variational degrees of freedom in the excited state. The overlap between the ground state and an excited state can be easily computed. We first introduce the following notation for a double-layer tensor containing a $B$ tensor in the ket-level,
\begin{equation}
\diagram{ansatz}{3} = \sum_{s,s'} \delta_{s,s'} B_{u,r,d,l}^s \bar{A}_{u',r',d',l'}^{s'}, 
\end{equation}
so that the overlap is given by
\begin{align} 
\braket{\Psi(A)|\Phi_{\vec{q}}(B)} &= 2\pi\delta(q_x)2\pi\delta(q_y) \diagram{ansatz}{4} \label{eq:overlapgs} \\
&= 2\pi\delta(q_x)2\pi\delta(q_y)  \diagram{ansatz}{5}\;.
\end{align}
The overlap with the ground state can be written as $\braket{\Psi(A)|\Phi_{\vec{q}}(B)} = 4\pi^2\delta^{(2)}(\vec{q}) \vect{g}\dag \vect{B}$, so that we confine the variational subspace to tensors that are orthogonal to $\vect{g}$. Here we have introduced bold-face notation for the vectorized version of a tensor.
\par As such, the variational subspace is ill defined because of the presence of zero modes. Indeed, the particular choice for the $B$ tensor
\begin{equation}
\diagram{ansatz}{6} = \e^{iq_x} \diagram{ansatz}{7} - \diagram{ansatz}{8},
\end{equation}
with $X$ a random matrix, yields a zero state
\begin{equation}
\braket{\Phi_{\vec{q}}(B)|\Phi_{\vec{q}}(B)} = 0.
\end{equation}
Equivalently, the choice
\begin{equation}
\diagram{ansatz}{6} = \e^{iq_y} \diagram{ansatz}{9} - \diagram{ansatz}{10},
\end{equation}
yields a zero mode.
\par In general the overlap between two states in our variational space can be written as
\begin{equation}
\braket{\Phi_{\vec{q}'}(B')|\Phi_{\vec{q}}(B)} = 2\pi\delta(q_x-q_x')2\pi\delta(q_y-q_y') N_{\vec{q}}(B,B'),
\end{equation}
and, similarly, for overlaps of the Hamiltonian we find
\begin{equation}
\bra{\Phi_{\vec{q}'}(B)}H\ket{\Phi_{\vec{q}}(B)} = 2\pi\delta(q_x-q_x')2\pi\delta(q_y-q_y') M_{\vec{q}}(B,B').
\end{equation}
In addition, since the parametrization of the states $\ket{\Phi(B)}$ is clearly linear in the tensor elements of $B$, we can rewrite these expressions as
\begin{align}
N_{\vec{q}}(B,B') &= (\vect{B}')\dag \vect{N}_{\vec{q}} \vect{B} \\
M_{\vec{q}}(B,B') &= (\vect{B}')\dag \vect{M}_{\vec{q}} \vect{B},
\end{align}
where $\vect{B}$ denotes the vectorized version of the tensor $B$, and $\vect{M}_{\vec{q}}$ and $\vect{N}_{\vec{q}}$ are ($D^2d$)-dimensional hermitian matrices.
\par The variational optimization of the excitation ansatz
\begin{equation}
\min_{B \in \mathcal{R} } \frac{\bra{\Phi_{\vec{q}}(B)}H\ket{\Phi_{\vec{q}}(B)}}{\braket{\Phi_{\vec{q}}(B)|\Phi_{\vec{q}}(B)}},
\end{equation}
can now be rephrased as a generalized eigenvalue problem
\begin{equation}
 \left( \vect{P}\dag \vect{H}_{\vec{q}} \vect{P} \right) \vect{x}_{\vec{q}} = \omega(\vec{q}) \left( \vect{P}\dag \vect{N}_{\vec{q}} \vect{P} \right) \vect{x}_{\vec{q}}, 
\end{equation}
where $\vect{P}$ is a reduced basis for the $B$ tensors where all zero modes have been eliminated.

\subsection{The effective norm matrix}

We compute the effective norm matrix, for which we have to compute the matrix elements
\begin{equation}
\braket{\Phi_{\vec{q}'}(B')|\Phi_{\vec{q}}(B)} = 2\pi\delta(q_x-q_x')2\pi\delta(q_y-q_y') N_{\vec{q}}(B,B').
\end{equation}
The expression for $N_{\vec{q}}$ consists of a sum over all relative orientations of $B$ and $B'$ tensors, i.e. diagrams of the form
\begin{equation}
\diagram{neff}{1},
\end{equation}
where we have introduced the tensor notation
\begin{equation}
\diagram{neff}{2} = \sum_{s,s'} \delta_{s,s'} A_{u,r,d,l}^s \bar{B'}_{u',r',d',l'}^{s'}.
\end{equation}
First we need to define the following tensor capturing the infinite sum of the $B$ tensor going in the upper channel
\begin{align}
\diagram{neff}{3} &= \diagram{neff}{4}, \qquad \diagram{neff}{5} = \diagram{neff}{6} \\
\diagram{neff}{7} &= \diagram{neff}{8}, \qquad \diagram{neff}{9} = \diagram{neff}{10}.
\end{align}
Here we have introduced the 'pseudo inverse' of a channel operator, which appears whenever we want to take an infinite sum of contributions along a channel. Indeed, if we want to represent
\begin{equation}
\diagram{neff}{11} = \e^{iq_y} \diagram{neff}{12} + \e^{i2q_y} \diagram{neff}{13} + \e^{i3q_y} \diagram{neff}{14} + \dots,
\end{equation}
we should be able to compute
\begin{equation}
\sum_{n=0}^{\infty} \e^{iq_y(n+1)} \bigg( \diagram{neff}{15} \bigg)^n.
\end{equation}
This series converges if the spectral radius of the 'channel operator' $T$,
\begin{equation}
T = \diagram{neff}{15},
\end{equation}
is smaller than one. But, as we have seen above, we have normalized the tensors $M_l$ and $M_r$ such that the leading eigenvalue of this operator is exactly one, and the corresponding fixed point is given by $\rho$. Therefore we can write
\begin{equation}
\sum_{n=0}^{\infty} \e^{iq_y(n+1)} \bigg( \diagram{neff}{15} \bigg)^n = \e^{iq_y} \sum_{n=0}^{\infty} \e^{iq_yn} \bigg( \diagram{neff}{16} \bigg) + \e^{iq_y} \left(1 - \e^{iq_y} \bigg( \diagram{neff}{15} \bigg) \right)^P,
\end{equation}
where the notation $(\dots)^P$ implies that we have projected out the fixed point subspace from the channel operator. The fixed-point projector contains a potential divergence, but, since we always make sure that Eq.~\eqref{eq:overlapgs} is zero, this contribution will always drop out. 
\par These tensors we can use to define the following auxiliary tensor
\begin{align}
\diagram{neff}{17} = \e^{-iq_x} \diagram{neff}{18} + \diagram{neff}{19}
\end{align}
and we define three other $\gamma$ tensors that are related to the one above by a simple rotation and an interchange of the momenta. The full matrix element is then given by
\begin{align}
N_{\vec{q}}(B,B') &= \diagram{neff}{20} + \e^{+iq_y} \diagram{neff}{21} \nonumber \\ 
& \qquad\qquad + \e^{+iq_x} \diagram{neff}{22} + \e^{-iq_y} \diagram{neff}{23} + \e^{-iq_x} \diagram{neff}{24} \\
&= \diagram{neff}{20} + \e^{+iq_y} \diagram{neff}{21} + \rots.
\end{align}
where three rotated versions of the last diagram should be added with the appropriate momentum factors.

\subsection{The effective energy matrix}

We want to compute the matrix elements
\begin{equation}
\bra{\Phi_{\vec{q}'}(B')} H \ket{\Phi_{\vec{q}}(B)} = 2\pi\delta(q_x-q_x')2\pi\delta(q_y-q_y') M_{\vec{q}}(B,B')
\end{equation}
The matrix elements $M$ have a number of different contributions.

\subsubsection*{Purely local contributions}

The local contributions are given by all orientations where both $B$ and $B'$ are located on top of $h$. We first locate $B'$ on the upper-left site of the $h$ plaquette, and $B$ can hop on the four different sites. The other orientations of $B'$ are then obtained by simply rotating the same diagrams, and inserting the correct momentum factors:
\begin{align}
M^{\text{local}}_{q_xq_y}(B,B') &= \diagram{local}{1} + \e^{iq_x} \diagram{local}{2} \nonumber \\ 
& \qquad + \e^{iq_x}\e^{-iq_y} \diagram{local}{3} + \e^{-iq_y} \diagram{local}{4} \nonumber \\
& \qquad + \rots. 
\end{align}

\subsubsection*{Disconnected sums}

The next series of contributions correspond to the situations where two of the three objects are on the same site, whereas the third one is disconnected. Let us therefore first define a few new tensors, which contain infinite sums of disconnected operators. The first is the sum of disconnected $B$ tensors, 
\begin{equation}
\diagram{disconnected}{1} = \diagram{disconnected}{2}.
\end{equation}
Secondly, we have the sum of disconnected $h$ operators
\begin{equation}
\diagram{disconnected}{3} = \diagram{disconnected}{4},
\end{equation}
We further define the following tensors
\begin{align}
\diagram{disconnected}{5} &= \e^{-iq_x} \diagram{disconnected}{6} +  \diagram{disconnected}{7} \\
\diagram{disconnected}{8} &= \e^{+iq_x} \diagram{disconnected}{9} +  \diagram{disconnected}{10}
\end{align}
and their two-site versions
\begin{align}
\diagram{disconnected}{11} &= \e^{-2iq_x} \diagram{disconnected}{12} \nonumber \\
& \hspace{2cm} + \e^{-iq_x} \diagram{disconnected}{13} + \diagram{disconnected}{14} \\
\diagram{disconnected}{15} &= \e^{+2iq_x} \diagram{disconnected}{16} \nonumber \\
& \hspace{2cm} + \e^{+iq_x} \diagram{disconnected}{17} + \diagram{disconnected}{18} .
\end{align}
The Hamiltonian versions of these tensors are
\begin{align}
\diagram{disconnected}{19} &= \diagram{disconnected}{20} + \diagram{disconnected}{21} \\ 
\diagram{disconnected}{22} &= \diagram{disconnected}{23} + \diagram{disconnected}{24}.
\end{align}
Here we have introduced the channel operators $T_l$ and $T_r$ as
\begin{equation}
T_l = \diagram{disconnected}{25}, \qquad T_r = \diagram{disconnected}{26}.
\end{equation}

\subsubsection*{Semi-local contributions}

We can use these tensors to add the contributions where $B'$ is on either $B$ or $h$, while the other is disconnected. First, we have the contributions where $h$ is disconnected; we need to sum all diagrams of the form
\begin{equation}
\diagram{semilocal}{1},
\end{equation}
where $h$ is located completely in the upper-left part of the lattice. This sum amounts to the two following diagrams,
\begin{equation}
M^{\text{sl},1}_{q_xq_y}(B,B') = \diagram{semilocal}{2} +  \diagram{semilocal}{3} + \rots,
\end{equation}
and we have three rotated versions corresponding to the orientations of $h$ in the other quarters of the lattice. 
\par Secondly, when $B$ is disconnected, we sum all possible locations of $B$, and $B'$ is located on the first site of $h$. All other diagrams are then related through rotations:
\begin{align}
& M^{\text{sl},2}_{q_xq_y}(B,B') = \nonumber \\
& \qquad \e^{iq_y} \left( \diagram{semilocal}{4} + \e^{iq_x} \diagram{semilocal}{5} \right) \nonumber \\
& \qquad + \e^{i2q_x} \left( \diagram{semilocal}{6}  \right) + \e^{-2iq_y} \left( \e^{iq_x} \diagram{semilocal}{7}   \right) \nonumber \\
& \qquad + \e^{-iq_x} \left( \e^{-iq_y} \diagram{semilocal}{8} + \diagram{semilocal}{9} \right) \nonumber \\
& \qquad + \rots.
\end{align}

\subsubsection*{Non-local contributions}

We now compute all contributions where we can divide the lattice into two halves, where $B$ and $h$ are above the bipartition and the $B'$ tensor is below the line. That means we will have diagrams of the form
\begin{equation}
\diagram{nonlocal0}{1}, \quad \diagram{nonlocal0}{2}.
\end{equation}
Do note that we have diagrams which allow for two different bipartitions, for example
\begin{equation}
\diagram{nonlocal0}{3}, \quad \diagram{nonlocal0}{4}.
\end{equation}
In the following we will only take the diagrams of the left, so that we don't count diagrams twice.
\par Define the auxiliary tensors
\begin{align}
 &\diagram{gamma1}{1}  =  \e^{+2iq_y} \diagram{gamma1}{2} + \e^{iq_y} \diagram{gamma1}{3} \nonumber \\
 & \qquad + \left( \e^{-2iq_x} \diagram{gamma1}{4} + \e^{-iq_x} \diagram{gamma1}{5} \right. \nonumber \\
 & \qquad \hspace{2cm} \left.  + \e^{-2iq_x} \diagram{gamma1}{6} + \diagram{gamma1}{7} \right) \nonumber \\
 & \qquad + \e^{-2iq_x} \diagram{gamma1}{8} + \e^{-iq_x} \diagram{gamma1}{9} + \diagram{gamma1}{10}
\end{align}
and
\begin{align}
& \diagram{gamma2}{1} =  \e^{+2iq_y} \diagram{gamma2}{2} \nonumber \\
& + \e^{iq_y} \left( \e^{-2iq_x} \diagram{gamma2}{3}  + \e^{-iq_x} \diagram{gamma2}{4} \right. \nonumber \\
&  \hspace{7cm} \left. + \diagram{gamma2}{5} \right) \\
\end{align}
such that we can define
\begin{equation}
\diagram{gamma3}{1} = \diagram{gamma3}{2}, \quad \diagram{gamma3}{3} = \diagram{gamma3}{4}.
\end{equation}
Now we are in a position to add all contributions that we can obtain by making a bipartition. We group the terms in four groups, according to the orientation of $h$. First we sum the contributions where $h$ is situated in the up-right corner to $B'$, i.e. in the region
\begin{equation}
\diagram{nonlocal2}{1},
\end{equation}
and $B$ can be anywhere. All terms where $B$ and $h$ are to the right of $B'$ are contained in the $\gamma$ tensor, so that we have the following diagrams:
\begin{align}
& \diagram{nonlocal2a}{1} = \e^{iq_x} \diagram{nonlocal2a}{2} + \e^{iq_x} \diagram{nonlocal2a}{3} \nonumber \\
& \qquad + \left( \e^{i2q_y} \diagram{nonlocal2a}{4} + \e^{iq_y} \diagram{nonlocal2a}{5}  + \diagram{nonlocal2a}{6} \right) \nonumber \\
& \qquad + \e^{-iq_x} \left( \e^{iq_y} \diagram{nonlocal2a}{7} + \diagram{nonlocal2a}{8} \right) .
\end{align}
Secondly, we have the terms where $h$ is right above $B'$ on the right side,
\begin{equation}
\diagram{nonlocal2}{2}
\end{equation}
and $B$ anywhere above $B'$, with the diagrams
\begin{align}
& \diagram{nonlocal2b}{1} = \e^{2iq_x} \left( \diagram{nonlocal2b}{2} + \diagram{nonlocal2b}{3} \right) \nonumber \\
& + \e^{iq_x} \left( \e^{i2q_y} \diagram{nonlocal2b}{4} + \e^{iq_y} \diagram{nonlocal2b}{5} \right. \nonumber \\
& \hspace{2cm} + \left. \diagram{nonlocal2b}{6} + \diagram{nonlocal2b}{7} \right)  \nonumber\\
& + \left( \e^{i2q_y} \diagram{nonlocal2b}{8} + \e^{iq_y} \diagram{nonlocal2b}{9} \right.  \nonumber \\
& \hspace{2cm} + \left. \diagram{nonlocal2b}{10} + \diagram{nonlocal2b}{11} \right) \nonumber \\
& + \e^{-iq_x} \left( \diagram{nonlocal2b}{12} + \diagram{nonlocal2b}{13} \right. \nonumber \\
& \hspace{2cm} \left. + \e^{iq_y} \diagram{nonlocal2b}{14} \right) .
\end{align}
Thirdly, we have $h$ in the region
\begin{equation}
\diagram{nonlocal2}{3},
\end{equation}
with the diagrams
\begin{align}
& \hspace{-1cm} \diagram{nonlocal2c}{1} = \nonumber \\
& \e^{iq_x} \left( \diagram{nonlocal2c}{2} + \diagram{nonlocal2c}{3} \right. \nonumber \\
& \hspace{2cm} + \left. \e^{iq_y} \diagram{nonlocal2c}{4} \right) \nonumber \\
& + \left( \e^{i2q_y} \diagram{nonlocal2c}{5} + \e^{iq_y} \diagram{nonlocal2c}{6} \right. \nonumber\\
& \hspace{2cm} + \left. \diagram{nonlocal2c}{7} + \diagram{nonlocal2c}{8} \right) \nonumber \\
& +  \e^{-iq_x} \left( \e^{i2q_y} \diagram{nonlocal2c}{9} + \e^{iq_y} \diagram{nonlocal2c}{10} \right. \nonumber \\
& \hspace{2cm} + \left. \diagram{nonlocal2c}{11} + \diagram{nonlocal2c}{12} \right) \nonumber \\
& + \e^{-i2q_x} \left( \diagram{nonlocal2c}{13} + \diagram{nonlocal2c}{14} \right).
\end{align}
The last contribution is given by the orientation of $h$ as given by
\begin{equation}
\diagram{nonlocal2}{4},
\end{equation}
where this time we have to make sure that we don't add the contributions that can be obtained by another bipartition:
\begin{multline}
\diagram{nonlocal2d}{1} = \e^{iq_x} \left( \e^{iq_y} \diagram{nonlocal2d}{2} + \diagram{nonlocal2d}{3} \right) \\
\left(  \e^{2iq_y} \diagram{nonlocal2d}{4} +  \e^{iq_y} \diagram{nonlocal2d}{5} + \diagram{nonlocal2d}{6} \right).
\end{multline}
\par Now we construct the tensor which adds these four contributions (with $\alpha_t=\sum_{i=1\dots4}\alpha_i$)
\begin{equation}
\diagram{nonlocal0}{5} = \diagram{nonlocal0}{6},
\end{equation}
and, finally
\begin{equation}
M_{\vec{q}}^{\text{nl}}(B,B') = \e^{iq_y} \diagram{nonlocal0}{7} + \rots.
\end{equation}

\subsubsection*{Special disconnected contributions}

In this subsection, we compute all diagrams which cannot be obtained by making a horizontal bipartition in the lattice where $B'$ is on the one side, and $B$ and $h$ are on the other. We compute all terms for which $h$ is in the first quarter as defined by $B'$, i.e. in the upper-left region
\begin{equation}
\diagram{special}{1},
\end{equation}
and $B$ can be anywhere such that there is no bipartitation possible. The three other orientations of $h$ are related through rotations of these diagrams.
\par All terms are given by
\begin{align}
& M_{\vec{q}}^{\text{sp}}(B,B') = \nonumber \\
& \qquad \e^{+iq_x} \diagram{special}{2} + \e^{-iq_y} \diagram{special}{3} \nonumber \\
& \qquad + \e^{-iq_y} \e^{-iq_x} \diagram{special}{4} + \e^{-2iq_x} \diagram{special}{5} \nonumber \\
& \hspace{5cm}  + \e^{-iq_x} \diagram{special}{6} \nonumber \\
& \qquad + \e^{-iq_y} \diagram{special}{7} + \e^{iq_x} \diagram{special}{8} + \e^{iq_x} \e^{iq_y} \diagram{special}{9} \nonumber \\
& \hspace{5cm}  + \e^{iq_y}\diagram{special}{10} + \e^{2iq_y}\diagram{special}{11}.
\end{align}

\end{document}